\documentclass[structabstract]{raa_twocloumn}
\usepackage{graphicx,times}
\usepackage{threeparttable}
\usepackage{natbib,amssymb}
\usepackage{float}
\usepackage[figuresright]{rotating}
\usepackage{amsmath}
\usepackage{supertabular}
\usepackage{multirow}
\begin{document}
   \title{The local and global properties of different types of supernova host galaxies}
   \volnopage{ {\bf 2019} Vol.\ {\bf XX} No. {\bf XXX}, 000--000}
   \setcounter{page}{1}
   \author{Li Zhou \inst{1,2},Yan-Chun Liang \inst{2}, Jun-Qiang Ge \inst{2}, Yi-Nan Zhu \inst{1}, Xu Shao \inst{2}, Hong Wu \inst{2}, Wei-Bin Shi \inst{3}, Li-Cai Deng \inst{2}}
   \institute{
   Department of Astronomy, School of Physics and Astronomy, Sun Yat-sen University, Zhuhai 519082, China; {\it zhouli59@mail.sysu.edu.cn}\\
   \and
   National Astronomical Observatories, Chinese Academy of Sciences,
         20A Datun Road, Chaoyang District, Beijing 100012, China;
         {\it ycliang@nao.cas.cn}\\
         \and
   Shandong Provincial Key Laboratory of Optical Astronomy and
  Solar-Terrestrial Environment, School of Space Science and Physics,
  Shandong University at Weihai, Weihai 264209, China }  
   \date{Received; accepted}

\abstract{By using Data Analysis Pipeline (DAP) products of Mapping Nearby Galaxies at Apache Point Observatory (MaNGA), which are publicly available from the SDSS Data Release 15, we analyze the local properties at the SN explosion sites and global properties of different types of SN host galaxies to  explore the explosion environments of different types of SNe. In our sample, there are 67 SN host galaxies in the field of view of MaNGA, including 32 Type Ia, 29 CCSNe, 1 super-luminous SN (SLSN), 1 Type I and 4 unclassified type of SNe, with which we can perform the K-S test for analysis and derive statistically robust results. Due to the limited sample size, we couldn't remove the mass dependence in this work, which is likely the true driver of the trends for the properties presented in this work. The global star formation rate (SFR) and EW(H$\alpha$) for SN Ia hosts is slightly lower than that for CCSN hosts on average. SN Ia host galaxies are $\sim$ 0.3 dex more massive than CCSN hosts, which implies that the number ratio of CCSNe to Type Ia SNe will decrease with the increasing of stellar mass of host galaxies. The stellar population age of SN Ia host galaxies is older than that of CCSN hosts on average. There is no significant difference between different types of SN hosts for some properties, including local SFR density ($\Sigma \rm SFR$), local and global gas-phase oxygen abundance. For most galaxies in our sample, the global gas-phase oxygen abundance estimated from the integrated spectra of SN hosts can represent the local gas-phase oxygen abundance at the SN explosion sites with small bias.    
 }
\authorrunning{Zhou et al.}
\titlerunning{The properties of different types of SN hosts}
\maketitle

\section{Introduction}
\label{sec:intro}
As one of the most violent and important processes, supernova (SN) explosions mark the death of a stellar's life. 
Supernovae (SNe) can be classified into several types according to their different spectral features \citep{fi97,ha02,tu03,ba79,2016ApJ...820...33R,sc90}. 

A star with the initial stellar mass less than 8 M${_\odot}$ will explode to a degenerate carbon-oxygen (CO) white dwarf, which can grow by accretting materials from its non-degenerate companion star. When its   mass increases to 1.4 M$_\odot$, a SN Ia is produced after a bright thermonuclear explosion \citep{bi80,hf60}. 
SNe Ia are considered as standard candle and used for calculating the cosmology distance. 
They have made great contributions on the discovery of dark matter and the accelerating expansion of the Universe \citep{ri98,pe99}.  
Once a star has its initial mass larger than 8 M$_\odot$, an explosion caused by the gravitational collapse will happen and leave a neutron star or black hole as the remnant  \citep{be79,ar89}. This Core-Collapse supernova (CCSN) include Type Ib, Ic and Type II SNe.

Many efforts have been devoted to study the correlations between different types of SNe and their host galaxies by using single fiber spectra \citep{ha10,ke12,pr08,sh14}. 
With the development of integral field of spectroscopy (IFS) on modern telescopes, the local environment of the SNe and their host galaxies can be obtained with the IFS observations of SN host galaxies \citep{st12,ku13a,ku13b,ga16a}.


Many works focusing on the SN host galaxies have performed statistical analysis on the local and global properties of different types of SNe host galaxies.
\citet{ga14} studied 95 different types of SNe hosted in 81 galaxies using the IFS of the Calar Alto Legacy
Integral Field Area (CALIFA), the PPAK IFS Nearby Galaxies Survey (PINGS) and some other observations.
\citet{ga16b} obtained a larger number of SN host galaxies (115) and analyzed their metallicity with IFS data. 
With both IFS and long-slit data, \citet{ly18} studied the environment of 37 SNe Iax (SN Iax is a peculiar SN class and differ from normal SNe Ia) explosion sites and their host galaxies.
\citet{ku18} also used the IFS data to analyze the properties of 83 nearby CCSN explosion sites.
Based on the IFS observations from PMAS/PPak Integral-field Supernova hosts COmpilation (PISCO), \citet{ga18} analyzed the local and global properties of 232 SN host galaxies, which hosted 272 SNe.  

There are several works analyzed the environment of SN explosion sites for a small sample using the Mapping Nearby Galaxies at APO (MaNGA, \citealt{bu15} ). With MaNGA IFU data, \citet{ch17} and \citet{iz18} have analyzed the local environment of SLSN 2017egm in detail, which is one of the most nearby superluminous supernova.  
\citet{2019RAA....19..121Z} selected 11 SNe host galaxies from MaNGA DR13 and provided detailed information of the local and global properties of SNe explosion sites and their host galaxies one by one. 

In this paper, we will enlarge the sample size and analyze the differences of the local and global properties for  different types of SN host galaxies (32 Type Ia, 29 CCSNe, 1 SLSN, 1 Type I and 4 unclassified types) by using their MaNGA IFS observations to derive a more statistically robust conclusion. Also, we will compare the gas-phase oxygen abundance in the region of SNe explosion sites with those of the global regions of the host galaxies. Most of our sample galaxies have redshifts around 0.03 following the sample selection of MaNGA, which are systematically higher than other samples established in previous works. 

The paper is
organized as follows: in Sect.~\ref{sample selection}, we introduce the sample selection method used to
select our SN host galaxies. We arrange the comparisons of the local and global properties of environment for different types of supernova host galaxies in Sect.~\ref{results}. Finally, we
will discuss our results in Sect.~\ref{discussion} and conclude in Sect.~\ref{conclusion}. Throughout this paper, we adopt
a cosmological model with $H_0$ = 70 $km s^{-1} Mpc ^{-1}$, $\Omega_M$ = 0.3, $\Omega_{\Lambda} $= 0.7.

\section{Sample selection and Data reduction}
\label{sample selection}
We cross-correlate $\sim$ 4600 galaxies in the SDSS DR15 \citep{2019ApJS..240...23A,2019AJ....158..231W} with Asiago Supernova Catalogue (ASC), Sternberg Astronomical Institute (SAI) supernova catalogue and supernovae in Transient Name Server (TNS) website. The detailed sample selection procedures are described as follows.

\subsection{Supernova Catalogue}
\label{asc}
With the development, ASC presents the information of more and more SNe and host galaxies through several years \citep{ba84,ba89,ba99}. Although the last input of ASC was 2017jmj, which was discovered on 2017 December 31, ASC will check and update the information of the listed SNe and their host galaxies. Detailed information of ASC is presented in  http://graspa.oapd.inaf.it/. 

The last modified version of SAI was on 2014 October 17 and there were 6545 SNe in this catalogue. See detailed information of SAI supernova catalogue in http://www.sai.msu.su/sn/sncat/.     

TNS is an official mechanism of IAU and it is used to report new astronomical transient events. Once a transient event is spectroscopically confirmed,  this new SN discovery will be officially named in the form of the naming rules established by IAU. Before spectroscopically classified, the transient event has a prefix of 'AT'. If it is classified into any type of SN, the prefix will be changed into 'SN'. There are 3967 classified SNe released by TNS in public from 2016 January 1 to 2019 August 31. See detailed information of TNS in 
https://wis-tns.weizmann.ac.il/.    

\subsection{MaNGA DAP data products}
\label{manga}
Mapping Nearby Galaxies at Apache Point Observatory (MaNGA) is one of three core programs of SDSS-IV and is designed to observe 10,000 galaxies with the integral field of spectroscopy. The survey covers galaxies with their redshift ranging from 0 to 0.15, and stellar mass from $\rm 10^9$ $\rm M_\odot$ to $\rm 10^{11}$ $\rm M_\odot$, which can help the clarification on the galaxy evolution processes from the birth, growth, and finally their death \citep{we15,bu15,la15,dr15,ba17}. 
MaNGA galaxies are divided into two kinds of samples based on their spatial coverage, i.e. the Primary sample for 2/3 and the Secondary sample for 1/3 of galaxies, which can cover galaxies out to 1.5 $R_e$ and 2.5 $R_e$, with a median redshift of 0.03 and 0.045, respectively (More details in \citealt{bu15,la15,wa17,ya16b}). 

The MaNGA Data Analysis Pipeline (DAP) was initially developed in 2014 with an original IDL version and used as a survey-level pipeline to provide data products to the SDSS collaboration. It has been evolved through several years to a python implementation and now is flexible to prospective users \citep{2019ApJS..240...23A, 2019AJ....158..231W}. The MaNGA DAP products provide galaxy parameters extracted from
both emission and absorption lines \citep{2019AJ....158..231W}, including spectral indexes, emission-line properties, and gas and stellar kinematics, etc.

\subsection{Cross-correlations of catalogs and final sample}
\label{sample}
To obtain SN host galaxies in the field of view (FoV) of MaNGA, we match the R.A. and Dec. of SNe in ASC, SAI supernova catalogue and TNS with the R.A and Dec. of more than 4600 galaxies in SDSS DR15. The cross-match radius is set to 15\arcsec  to match the largest FoV of MaNGA. Then we draw the 2D maps and exclude the galaxies that the  SNe locate out of the FoV of MaNGA. Finally, we select 67 SN host galaxies in this matching radius that have been observed.
Table~\ref{table.allsamples-67} presents detailed information. The '+' or '-' symbols in the table represent there are AGNs or there are no AGNs in the galaxy centers according to BPT diagram and more details will be described in sec~\ref{bpt}. Through this table, we can see that there are 32 Type Ia SNe, 29 CCSNe, 1 SLSN (2017egm), 1 Type I SN and 4 unclassified SNe.   
\begin{table*}
\begin{center}
\bottomcaption{Basic information from MaNGA and ASC for our 67 sample galaxies. The symbols of '+' represent that there is AGN in the galaxy nucleus and '-' represent that there is no AGN in the galaxy nucleus.}
\label{table.allsamples-67}
\tiny
\begin{supertabular}{|p{11mm}|p{15mm}|p{7mm}|p{10mm}|p{12mm}|p{26mm}|p{10mm}|p{10mm}|p{10mm}|p{10mm}|p{3mm}|}
Plateifu	&	SN Name	&	SN type	&	SN ra	&	SN dec	&	Host Name	&	Redshift	&	E(B-V)	&	NSA$_{b/a}$	&	NSA$_{PA}$ & AGN\\
           &           &           & $[deg]$   & $[deg]$   &             &               &    $[mag]$  &            &              &   \\
\hline	
8084-12702	&	2019abu	&	II	&	50.536613	&	-0.84006111	&		&	0.0364594	&	0.0656572	&	0.860265	&	47.9614	&	+	\\
9871-12702	&	2018lev	&	II	&	228.28404	&	41.2637	&		&	0.0290635	&	0.0214573	&	0.545892	&	160.221	&	+	\\
8080-12703	&	2018jfp	&	II	&	49.484488	&	-0.16970556	&		&	0.0227967	&	0.0727471	&	0.704163	&	61.6702	&	+	\\
8257-12701	&	2018hfc	&	IIP	&	165.49408	&	45.227539	&		&	0.0199963	&	0.00869303	&	0.580468	&	72.7335	&	-	\\
9872-9102	&	2018fbh	&	Ia	&	234.13327	&	41.799897	&		&	0.0413954	&	0.0331264	&	0.808343	&	120.581	&	+	\\
8262-6104	&	2018ddh	&	Ia	&	184.68351	&	44.781975	&		&	0.0383413	&	0.0124629	&	0.620135	&	121.685	&	+	\\
8602-12701	&	2018ccl	&	Ia	&	247.04635	&	39.820131	&		&	0.0267882	&	0.00939909	&	0.649031	&	2.06161	&	+	\\
8945-9102	&	2018btb	&	Ia	&	173.61624	&	46.362531	&		&	0.0338424	&	0.0182859	&	0.839694	&	115.388	&	-	\\
8611-6101	&	2018bbz	&	Ia	&	261.96804	&	60.096056	&	PGC60339	&	0.0278347	&	0.0259684	&	0.840635	&	59.5055	&	+	\\
8448-6102	&	2018aex	&	II	&	165.18908	&	22.2875	&		&	0.0229026	&	0.0173848	&	0.223525	&	117.637	&	-	\\
8315-12704	&	2018aej	&	Ia	&	236.09589	&	39.558094	&		&	0.0479339	&	0.0165416	&	0.255403	&	175.803	&	-	\\
8083-12704	&	2017jcu	&	II	&	50.695417	&	0.14808333	&	UGC02705	&	0.0228308	&	0.124875	&	0.894133	&	154.66	&	+	\\
8147-12701	&	2017ixz	&	IIb	&	116.76261	&	26.773825	&		&	0.0235924	&	0.0377963	&	0.649183	&	112.922	&	-	\\
9029-6102	&	2017frc	&	II	&	246.39147	&	41.682189	&		&	0.0279312	&	0.00923756	&	0.968164	&	68.0598	&	-	\\
8616-12702	&	2017fel	&	Ia	&	322.30573	&	-0.29470278	&		&	0.0305366	&	0.0524228	&	0.775617	&	76.7474	&	+	\\
9095-6102	&	2017ets	&	Ia	&	243.44229	&	22.916428	&		&	0.0319161	&	0.0903559	&	0.684788	&	161.49	&	-	\\
8454-12703	&	2017egm	&	SLSN-I	&	154.77342	&	46.453911	&		&	0.0307213	&	0.0113688	&	0.796432	&	5.52863	&	-	\\
9872-12703	&	2017dit	&	Ia	&	231.9955	&	42.846861	&		&	0.0185896	&	0.026274	&	0.808231	&	100.692	&	-	\\
9041-12702	&	2017dgs	&	II	&	236.38983	&	30.145431	&		&	0.0316281	&	0.0331679	&	0.406954	&	166.074	&	-	\\
9043-12701	&	2017def	&	Ia	&	230.62042	&	27.707189	&		&	0.0748986	&	0.0407621	&	0.633883	&	164.034	&	+	\\
8625-9101	&	2017cxz	&	Ib	&	259.83242	&	57.898833	&	PGC140771	&	0.0289486	&	0.031763	&	0.414263	&	7.11978	&	-	\\
8145-1902	&	2017ckx	&	Ia	&	117.04586	&	28.230289	&	UGC 04030 NED01	&	0.0271584	&	0.0341972	&	0.379181	&	81.3	&	-	\\
8149-9102	&	2017cfq	&	II   &	120.98008	&	26.520167	&	IC 0491	&	0.0217485	&	0.0427882	&	0.436642	&	114.077	&	-	\\
8547-12701	&	2017boa	&	Ia	&	217.63179	&	52.704325	&	SDSS J143031.19+524225.8	&	0.0448811	&	0.0114932	&	0.47447	&	164.296	&	+	\\
9501-12705	&	2016gkm	&	Ib/c	&	130.72108	&	25.070919	&		&	0.0172569	&	0.032717	&	0.866949	&	136.132	&	-	\\
9509-12703	&	2016acq	&	II	&	124.10248	&	25.993122	&		&	0.0452423	&	0.0354447	&	0.728145	&	127.374	&	+	\\
8550-9101	&	2015cq	&	II	&	247.41083	&	40.236111	&	SDSS J162938.32+401407.4	&	0.0283296	&	0.00875131	&	0.177729	&	74.261	&	-	\\
7960-12705	&	2015co	&	II	&	258.6505	&	30.735667	&	SDSS J171436.05+304400.7	&	0.0295939	&	0.0486122	&	0.544136	&	158.44	&	-	\\
9027-12701	&	2012fj	&	IIP	&	243.93692	&	31.96325	&	NGC 6103	&	0.0315467	&	0.0229677	&	0.693064	&	74.9238	&	-	\\
8453-12702	&	2012al	&	IIn	&	151.54837	&	47.294611	&	Anon.	&	0.0381068	&	0.0111072	&	0.474214	&	167.375	&	-	\\
8588-6101	&	2011cc	&	IIn	&	248.456	&	39.263531	&	IC 4612	&	0.0317611	&	0.0103404	&	0.954619	&	153.413	&	-	\\
7495-12702	&	2010ee	&	II	&	205.07496	&	26.353311	&	UGC 8652	&	0.0284013	&	0.0135025	&	0.189334	&	13.9973	&	-	\\
7968-9102	&	2010dl	&	Ia	&	323.75404	&	-0.51328056	&	IC 1391	&	0.030156	&	0.0382006	&	0.750483	&	77.4445	&	+	\\
9876-12703	&	2009L	&	Ia	&	194.70042	&	27.673781	&	NGC 4854	&	0.0279622	&	0.010924	&	0.76496	&	53.2682	&	+	\\
8261-12705	&	2007sw	&	Ia	&	183.40367	&	46.493361	&	UGC 7228	&	0.0257328	&	0.0187694	&	0.332757	&	179.967	&	-	\\
8080-12705	&	2007gl	&	Ib/c	&	47.888371	&	-0.74631111	&	Anon.	&	0.0282305	&	0.0645811	&	0.922932	&	11.7656	&	-	\\
8138-12704	&	2007R	&	Ia	&	116.65637	&	44.789469	&	UGC 4008	&	0.030805	&	0.046767	&	0.625464	&	165.866	&	-	\\
9033-12705	&	2007O	&	Ia	&	224.02158	&	45.404689	&	UGC 9612	&	0.0361875	&	0.0231061	&	0.694846	&	89.3529	&	-	\\
9193-12704	&	2006np	&	Ia	&	46.6645	&	0.064030556	&	Anon.	&	0.107438	&	0.083128	&	0.66192	&	24.9588	&	+	\\
7975-6104	&	2006iq	&	Ia	&	324.89062	&	10.484861	&	Anon.	&	0.0788547	&	0.0610815	&	0.87799	&	98.1721	&	-	\\
8322-12705	&	2006cq	&	Ia	&	201.10458	&	30.956311	&	IC 4239	&	0.0485022	&	0.0150732	&	0.732988	&	141.256	&	+	\\
9036-9102	&	2005bk	&	Ic	&	240.571	&	42.915361	&	MCG +07-33-27	&	0.0244346	&	0.0147117	&	0.950868	&	146.291	&	-	\\
8153-12702	&	2004hx	&	II	&	40.3025	&	-0.87943889	&	Anon.	&	0.0381235	&	0.0338392	&	0.474935	&	71.3878	&	-	\\
7990-3703	&	2004eb	&	II	&	262.10125	&	57.546031	&	NGC 6387	&	0.0285483	&	0.0385512	&	0.555031	&	86.9226	&	-	\\
8078-12701	&	2004I	&	II	&	40.880129	&	0.30853056	&	NGC 1072	&	0.0266695	&	0.037467	&	0.381776	&	11.3952	&	+	\\
8990-12702	&	2004H	&	Ia	&	173.49904	&	49.062919	&	IC 708	&	0.0316278	&	0.0195327	&	0.777672	&	168.268	&	+	\\
8984-9101	&	2003an	&	Ia	&	201.97312	&	28.508111	&	MCG +05-32-22	&	0.0369943	&	0.0134218	&	0.791246	&	32.2228	&	+	\\
8615-9102	&	2002ik	&	IIP	&	321.47613	&	0.41488889	&	Anon.	&	0.0317398	&	0.0545153	&	0.465725	&	82.503	&	-	\\
9027-9101	&	2002ci	&	Ia	&	243.90812	&	31.3215	&	UGC 10301	&	0.0221977	&	0.0332517	&	0.247826	&	12.5372	&	-	\\
9029-12702	&	2002cg	&	Ic	&	247.252	&	41.283439	&	UGC 10415	&	0.0318872	&	0.0112731	&	0.793745	&	80.5852	&	-	\\
8978-6102	&	2002aw	&	Ia	&	249.37108	&	40.880639	&	Anon.	&	0.0263847	&	0.00693859	&	0.242865	&	79.4155	&	+	\\
8323-6101	&	2002G	&	Ia	&	196.98025	&	34.085139	&	Anon.	&	0.03365	&	0.0126278	&	0.827951	&	60.7429	&	+	\\
8604-12701	&	2000cs	&	II	&	245.88433	&	39.12475	&	MCG +07-34-15	&	0.0350385	&	0.0088737	&	0.973792	&	166.75	&	+	\\
8588-12702	&	1991L	&	Ib/c	&	250.31262	&	39.292389	&	MCG +07-34-134	&	0.0305423	&	0.0144956	&	0.667299	&	79.4933	&	-	\\
8250-12704	&	1999 gw	&	Unknown &	138.97792	&	44.331944	&	UGC 4881	&	0.0397859	&	0.0171982	&	0.551148	&	67.5947	&	-	\\
8322-3701	&	1996 ce	&	Unknown &	199.06446	&	30.264611	&	Mrk785	&	0.0491864	&	0.0112192	&	0.803095	&	152.056	&	-	\\
8332-1902	&	2005 cc	&	Ia pec: &	209.27021	&	41.844944	&	NGC5383	&	0.00814191	&	0.00652679	&	0.704969	&	116.136	&	-	\\
8935-6104	&	1991 Q	&	Unknown	&	195.53163	&	27.648819	&	NGC4926A	&	0.023012	&	0.00914655	&	0.490692	&	74.2748	&	-	\\
9041-9102	&	2004 ct	&	Ia	&	235.94063	&	28.416528	&	M+05-37-17	&	0.0327509	&	0.0312202	&	0.791074	&	125.17	&	-	\\
9044-6101	&	1962 B	&	I	&	230.68879	&	29.768944	&	M+05-36-25	&	0.0229164	&	0.0267391	&	0.759202	&	92.9868	&	-	\\
8158-1901	&	ASASSN-15pn	&	Ia	&	60.85966667	&	-5.492027778	&	SDSS J040326.23-052930.6	&	0.0383575	&	0.0942628	&	0.489294	&	68.1569	&	-	\\
8311-6104	&	Gaia15abd	&	Ia	&	205.2829583	&	23.283	&		&	0.0263526	&	0.01702	&	0.783214	&	161.689	&	+	\\
8588-6102	&	ASASSN-15ns	&	Ia	&	250.11675	&	39.32021389	&	MRK887	&	0.0300793	&	0.013048	&	0.722699	&	17.4751	&	+	\\
9043-12704	&	ASASSN-15mm	&	II	&	231.3479167	&	29.17401944	&	SDSS J152523.40+291018.8	&	0.0214421	&	0.0237239	&	0.457505	&	1.30472	&	-	\\
9510-9101	&	ASASSN-15uv	&	Ia	&	126.783875	&	27.81159444	&	2MASX J08270817+2748382	&	0.0202794	&	0.0332955	&	0.402262	&	13.8534	&	-	\\
7815-3702	&	2017frb	&	Ia	&	317.9034583	&	11.49724444	&	CGCG426-012	&	0.0293823	&	0.101375	&	0.853027	&	170.98	&	-	\\
8550-12705	&	1975K	&	Unknown	&	249.13713	&	39.030139	&	NGC6195	&	0.029986	&	0.0185899	&	0.597257	&	38.9863	&	+	\\

\hline
\end{supertabular}
\end{center}
\end{table*}

\begin{figure}
\centering
\includegraphics[angle=0,width=7.8cm]{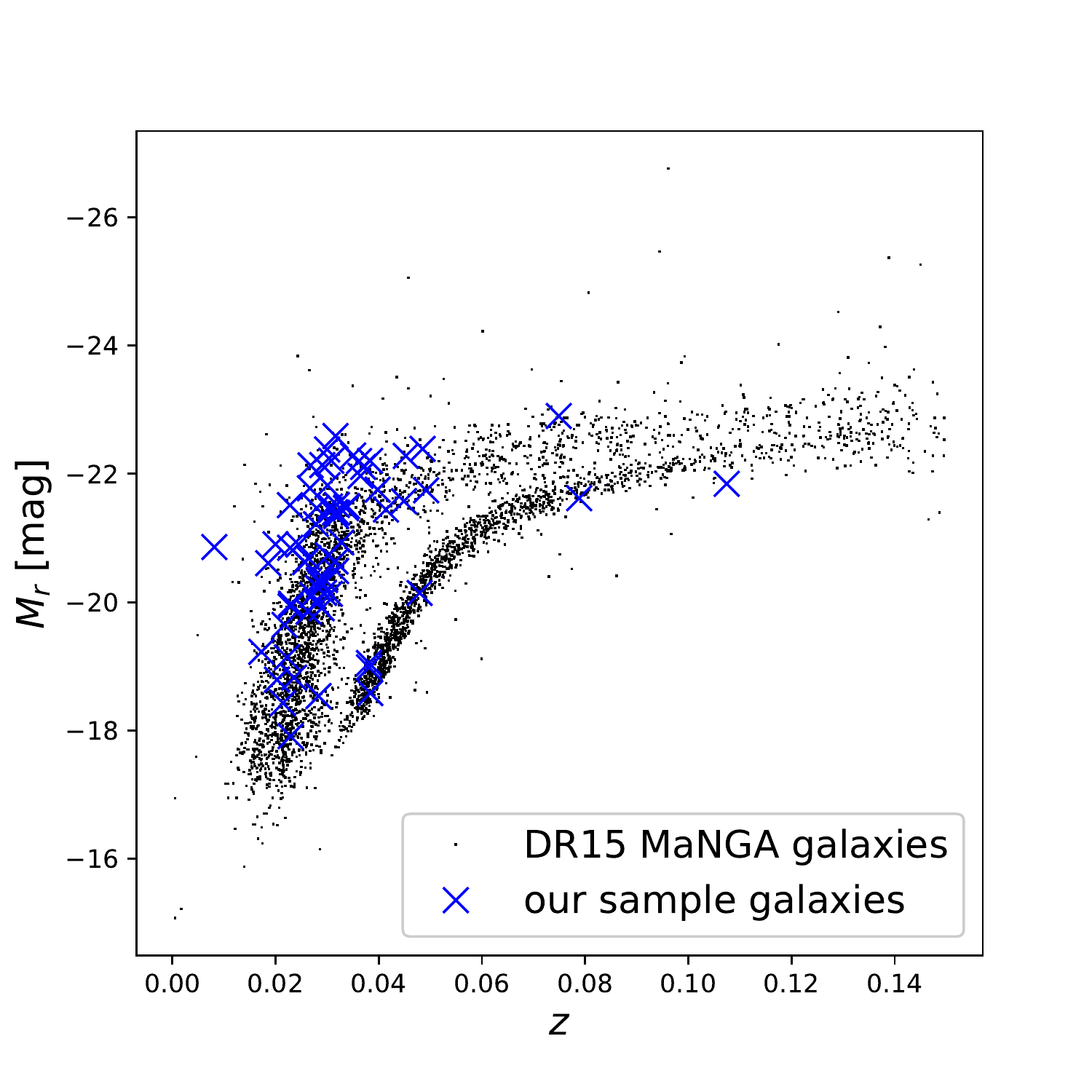}
\caption {The redshift and absolute magnitude in $r$ band distributions for all the 67 sample galaxies comparing with the DR15 MaNGA sample galaxies.}
\label{fig.z-Mr}
\end{figure}

\begin{figure}
\centering
\includegraphics[angle=0,width=7.8cm]{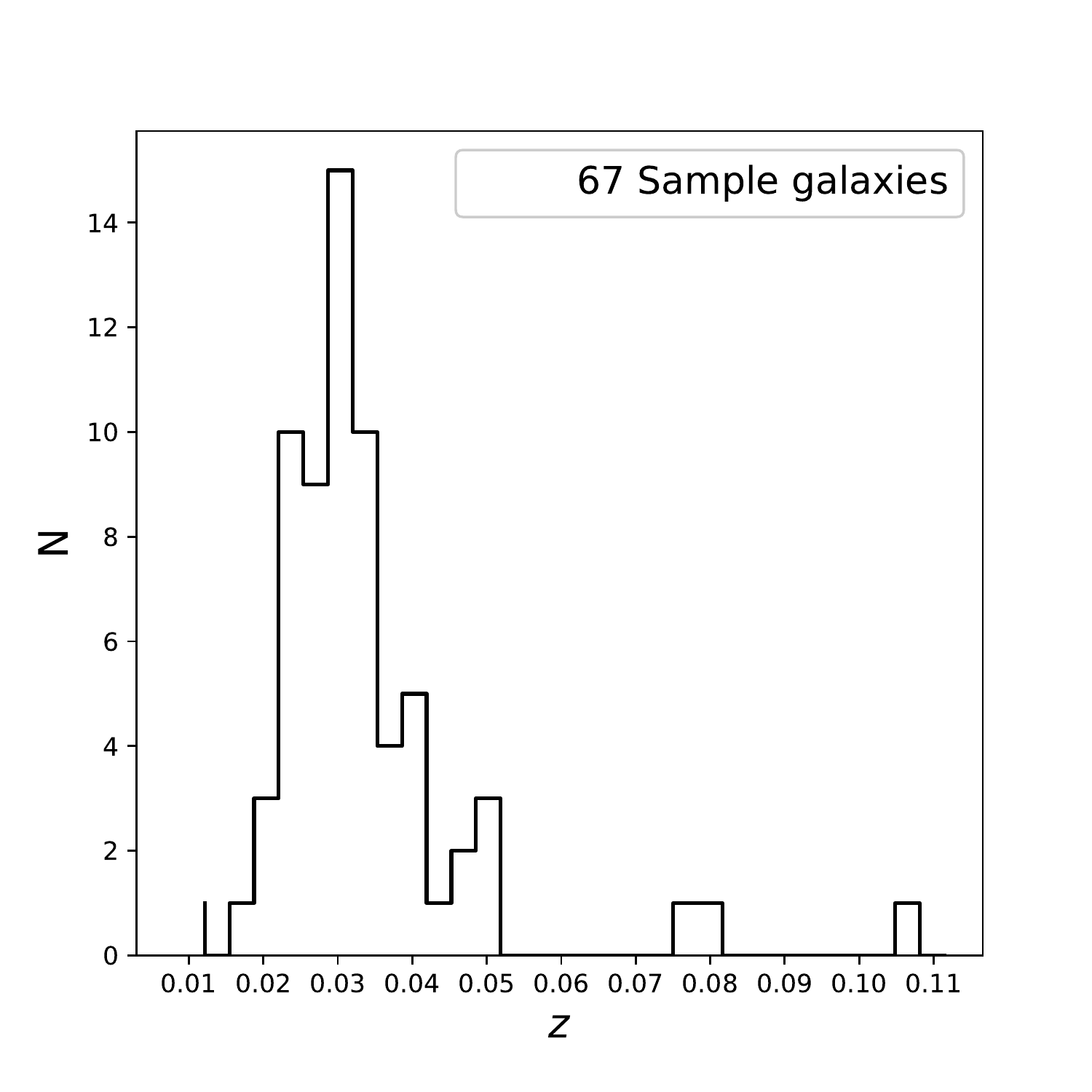}
\caption {The redshift distributions for all the 67 sample galaxies.}
\label{fig.z}
\end{figure}

Figure~\ref{fig.z-Mr} shows the range of redshift ($z$) and absolute magnitude in the $r$ band ($M_r$, which is taken from MaNGA DRP) for our 67 sample galaxies marked by crosses  and DR15 MaNGA galaxies marked by grey dots. The gap in this figure is resulted from the sample selection of MaNGA, which includes Primary and Secondary Samples \citep{wa15,be16}. From this figure, we can see that the majority of SN host galaxies come from the Primary Sample, and very few come from the Secondary Sample. There are two possible reasons that may cause this effect. The first possible reason is the number ratio of Primary and Secondary Sample of MaNGA. The Primary Sample is for 2/3 and the Secondary sample for 1/3 of MaNGA galaxies. 
The second possible reason is the range of the redshift of SN host galaxies. The median redshift for observed SNe is about 0.03. While the median redshifts of the Primary Sample and Secondary Sample are 0.03 and 0.045, respectively.  Therefore, we may obtain more galaxies from the Primary Sample after cross-correlating with SN catalogue.  According to this figure, the redshifts of our sample galaxies range from 0.01 to 0.12 with a median value about 0.03, and the $M_r$ of all the 67 sample galaxies range from -23 to -18 mag. From the histogram in Figure~\ref{fig.z}, we can see that the redshifts of most sample galaxies locate around 0.03. Table~\ref{table.allsamples-67} presents the  details of the redshift of each sample galaxy.    

\subsection{Parameter estimations}
\label{data reduction}
In our work, we use the emission line fluxes from MaNGA DAP to obtain the properties of dust extinction,  star-formation rate (SFR) and gas-phase oxygen abundance. The methods of calculating dust extinction, SFR and gas-phase oxygen abundance in this work are the same as \citet{2019RAA....19..121Z}. See  \citet{2019RAA....19..121Z} for more details.

We estimate the dust extinction using H$\alpha$ and H$\beta$ emission line ratios \citep{of06} and the equation of Rv=Av/E(B-V) \citep{fi99}. Then we do dust extinction correction for the emission line fluxes in the line of sight direction through galaxies.

According to \citet{ke98}, we estimate the ongoing SFR from H$\alpha$ luminosity. We calculate the global SFR by summing the SFR of all the single spaxels in 1.5 $R_e$ of galaxies. 

There are several methods to estimate the gas-phase oxygen abundance\citep{st06,iz06,li06,li07,yi07}. Here we use strong line methods of O3N2\citep{al79,pp04}, N2O2\citep{li06,do00,do13,2017MNRAS.466.3217Z} and R23\citep{pt05,pa79,mc91,za94,tr04,kd02,kk04,pi01} to obtain the global and local gas-phase oxygen abundance of SN host galaxies.

\section{Results}
\label{results}
The global properties are estimated using emission line fluxes from MaNGA DAP of all the useful spaxels in 1.5 $R_e$ of galaxies. The central and SN location properties are estimated using emission line fluxes of the spaxels within a circular region of 2.5 $arcsec$ radius around the galaxy centers and SN explosion positions. The global SFR and gas-phase oxygen abundance for galaxies without AGNs in the galaxy centers are estimated using emission line fluxes of spaxels in 1.5 $R_e$ of galaxies. While for galaxies with AGNs in the galaxy centers, the global SFR and gas-phase oxygen abundance are estimated using emission line fluxes of spaxels within 1.5 $R_e$ after excluding spaxels in central regions. In this section, firstly we will present the locations of galaxy centers in Baldwin, Phillips \& Terlevich (BPT) diagnostic diagram in Figure~\ref{fig.BPT} to check if there are active galactic nuclei (AGN) components in galaxy centers. Then we will compare the global properties of the host galaxies of SNe Ia and CCSNe and the results will be shown from Figure~\ref{fig.sfr-global} to Figure~\ref{fig.d4000-hdelta-67}. Also, we will compare the local galaxy properties at the SN explosion locations and show the results in Figure~\ref{fig.cdf-sfr-local-67} and Figure~\ref{fig.cdf-oh-local-67}. 
Table~\ref{table.med} shows the median and mean values of the global and local properties for the SN host galaxies in our sample.
\subsection{BPT diagnostic diagram}
\label{bpt}
\begin{figure}
\centering
\includegraphics[angle=0,width=7.8cm]{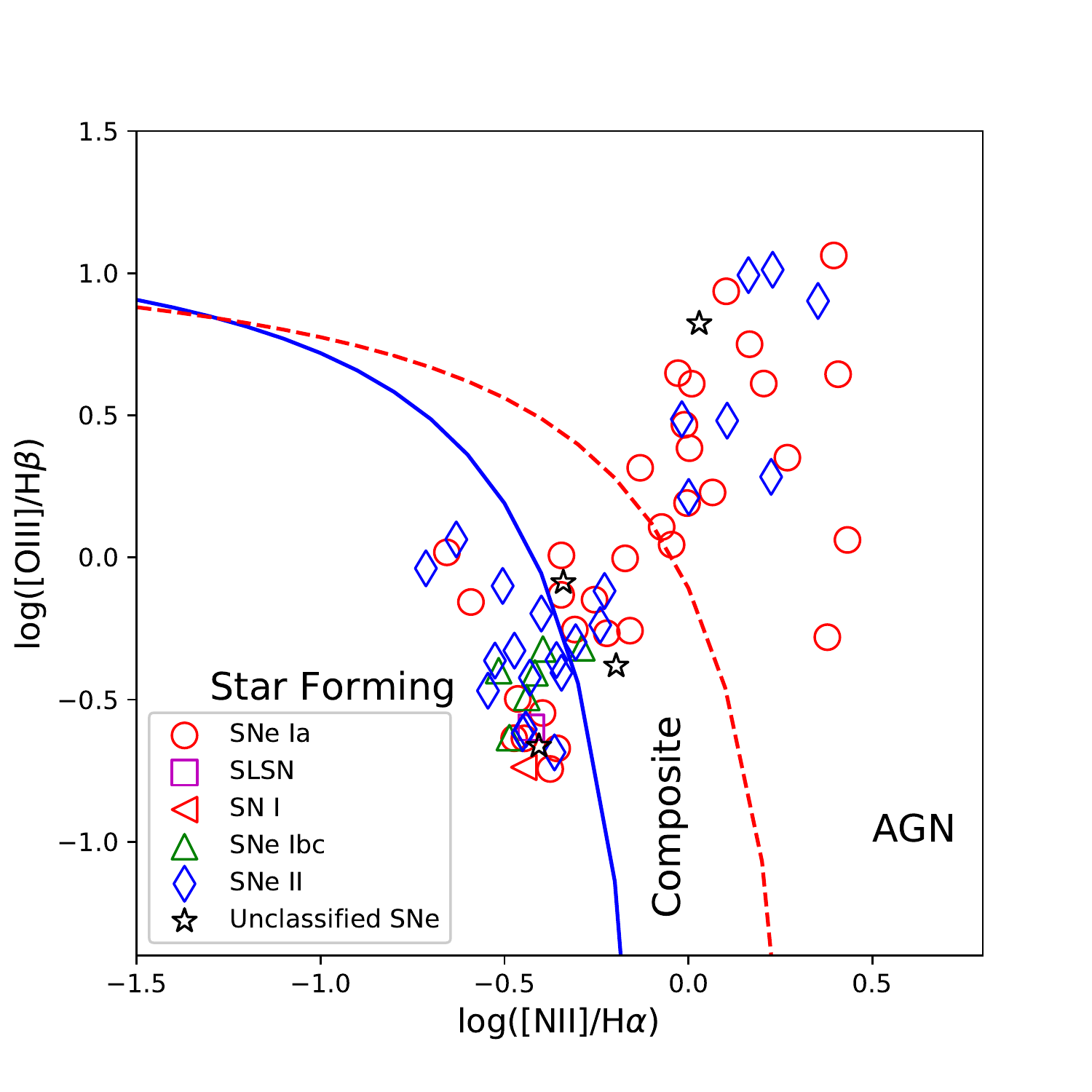}
\caption {The BPT diagram for our sample galaxies. We plot the emission-line flux ratio of log([O~{\sc iii}]/H$\beta$) versus the ratio log([N~{\sc ii}]/H$\alpha$) for the central regions of all the galaxies in our sample. Different symbols represent different types of SNe.}
\label{fig.BPT}
\end{figure}

To some extent, the existence of AGNs will bias the measurements. For galaxies with AGNs in the centers, the central regions should be excluded when calculating the global SFR and gas-phase oxygen abundance. Due to different excitation mechanism, AGNs can be well distinguished from star forming galaxies by the BPT diagram, which shows the flux ratio distribution of log([N~$\rm {\uppercase\expandafter{\romannumeral2}}$] $\lambda$6583/H$\rm \alpha$) in the horizontal axis and log([O~$\rm {\uppercase\expandafter{\romannumeral3}}$] $\lambda$5007/H$\beta$) in the vertical axis \citep{ba81}. 

Figure~\ref{fig.BPT} presents the locations of the central regions of our sample galaxies in the BPT diagram. In this figure, different symbols represent different types of SNe.
According to Figure~\ref{fig.BPT}, there are 25 sample galaxies with AGNs in the center. In Table~\ref{table.allsamples-67}, we use '+' to represent that there are AGNs in the center of a galaxy and '-' to represent a galaxy without AGNs. See details in Table~\ref{table.allsamples-67}. The number ratios of Type Ia, CCSNe and unclassified types of SNe host galaxies with AGNs in the center are 0.53 (17/32), 0.24 (7/29) and 0.25 (1/4), respectively. To reduce the impact of AGNs on global properties, we exclude the central regions of these 25 sample galaxies when estimating the global SFR and gas-phase oxygen abundance. 

\subsection{The difference of global properties between different types of SNe host galaxies}

\begin{figure}
\centering
\includegraphics[angle=0,width=7.8cm]{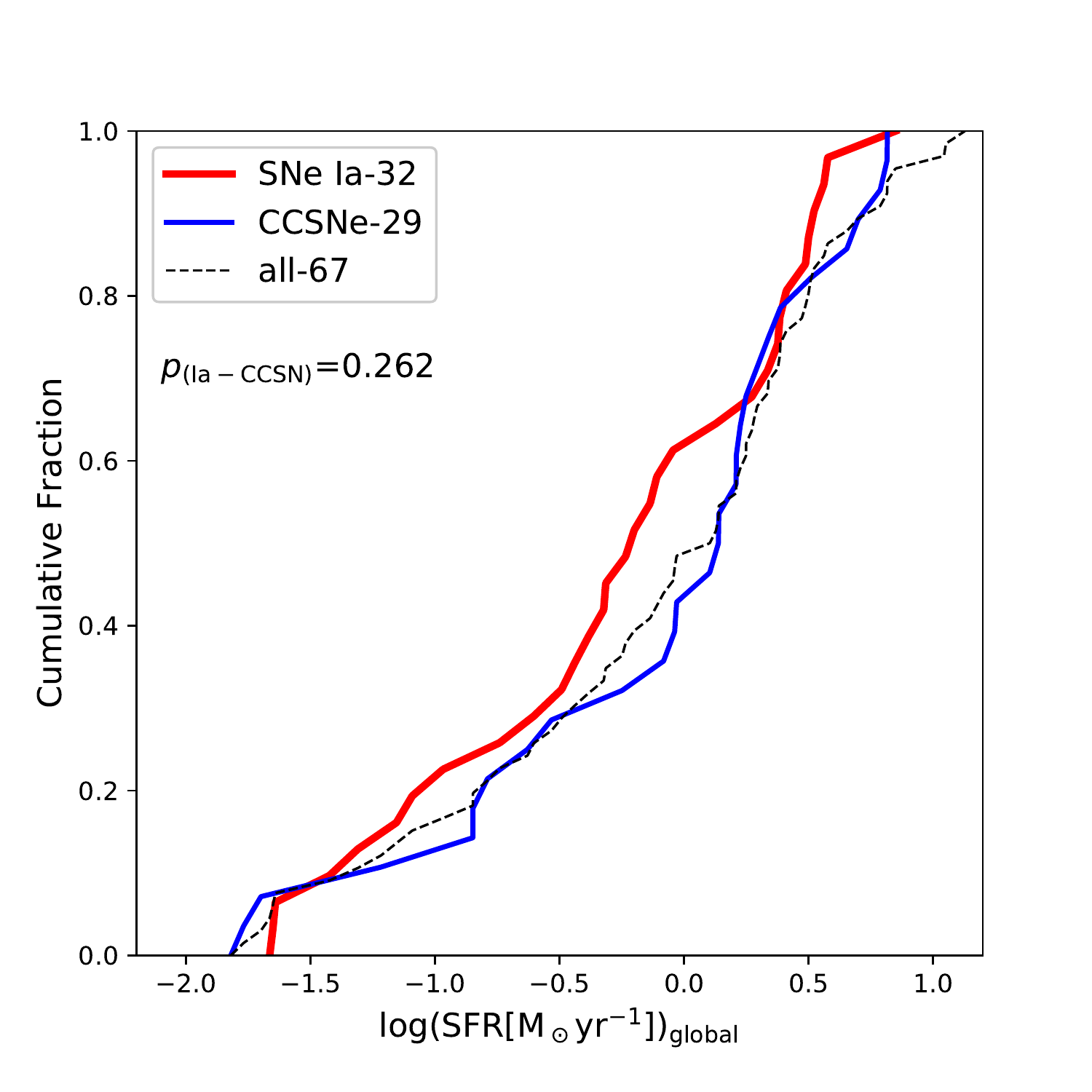}
\caption {The cumulative distributions of global SFR for different types of SN host galaxies. }
\label{fig.sfr-global}
\end{figure}

We estimate the global SFR through $H\alpha$ flux by summing the SFR of each useful spaxel in SNe host galaxies. Figure~\ref{fig.sfr-global} shows the cumulative distribution of global SFR for Type Ia SNe, CCSNe and all the sample galaxies, which are marked with red thick solid lines, blue thin solid lines and black dashed lines, respectively. We perform a Kolmogorov-Smirnov (K-S) test between SNe Ia and CCSNe host galaxies. According to the p-value of K-S test, which is 0.262, we can see that there is a difference between global SFR distribution of SNe Ia and that of CCSNe host galaxies. SNe Ia can explode in both late-type star forming galaxies and early-type galaxies, while CCSNe only explode in star forming galaxies. We can see from Figure~\ref{fig.sfr-global} and Table~\ref{table.med} that the average SFR of CCSN host galaxies is higher than that of SN Ia host galaxies, which means that CCSN host galaxies have stronger star forming activities. 

\begin{figure}
\centering
\includegraphics[angle=0,width=7.8cm]{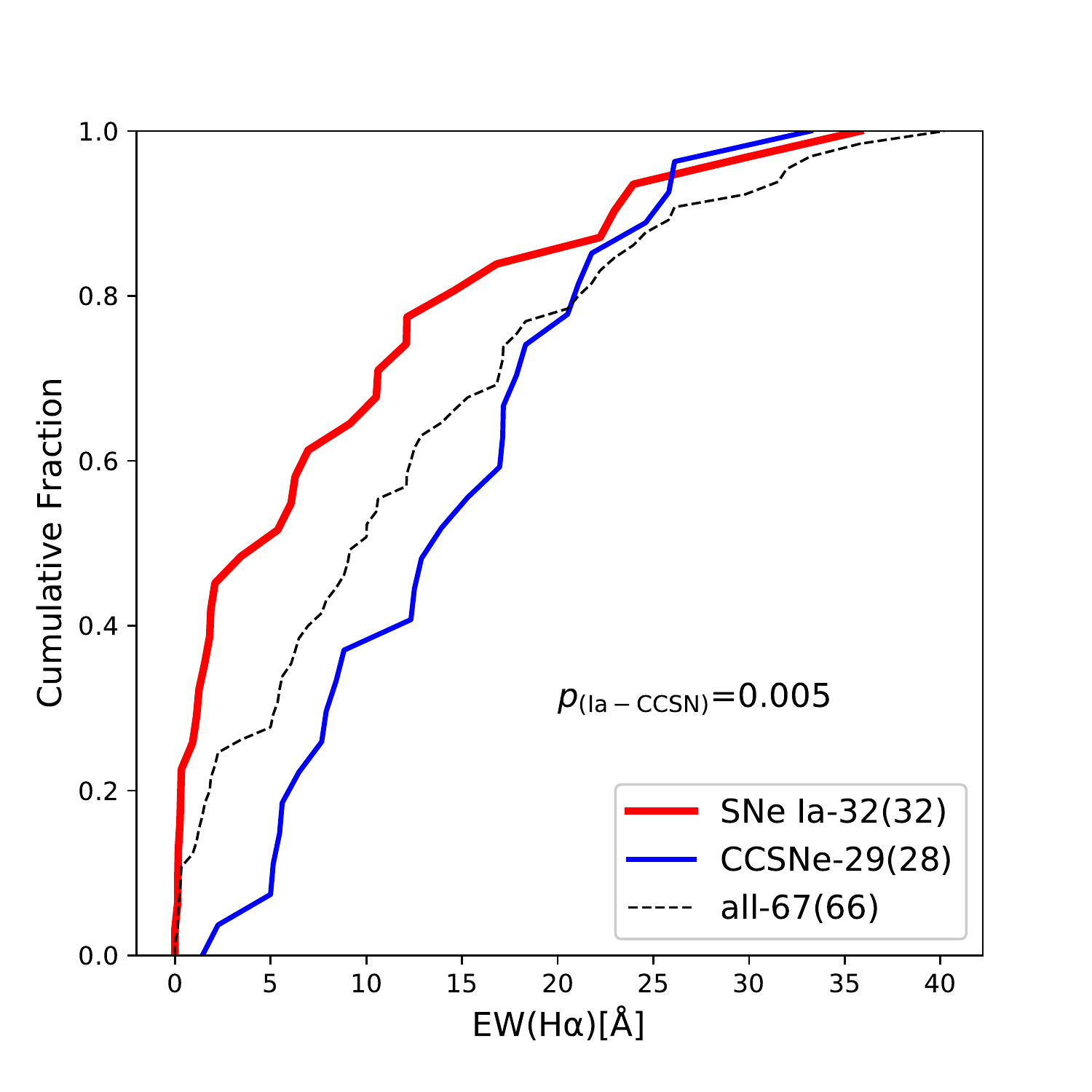}
\caption {The cumulative distributions of global EW(H$\alpha$) for different types of SN host galaxies. }
\label{fig.haew-global}
\end{figure}

The equivalent width of $H\alpha$ (EW($H\alpha$)) is thought to be an indicator of the strength of ongoing SFR compared with the past SFR \citep{2013A&A...554A..58S,ga14}. In this work, the global EW($H\alpha$) is estimated by calculating the median value of the spaxels in 1.5 $R_e$ of galaxies.
We present the cumulative distributions of the global EW(H$\alpha$) for SN Ia and CCSN host galaxies in Figure~\ref{fig.haew-global}. According to the p-value of K-S test, there is a significant difference between the global EW($H\alpha$) of these two types of SN host galaxies. From Figure~\ref{fig.haew-global} and Table~\ref{table.med}, we can see that the EW(H$\alpha$) of CCSN host galaxies is higher than that of SN Ia hosts. 

\begin{figure}
\centering
\includegraphics[angle=0,width=7.8cm]{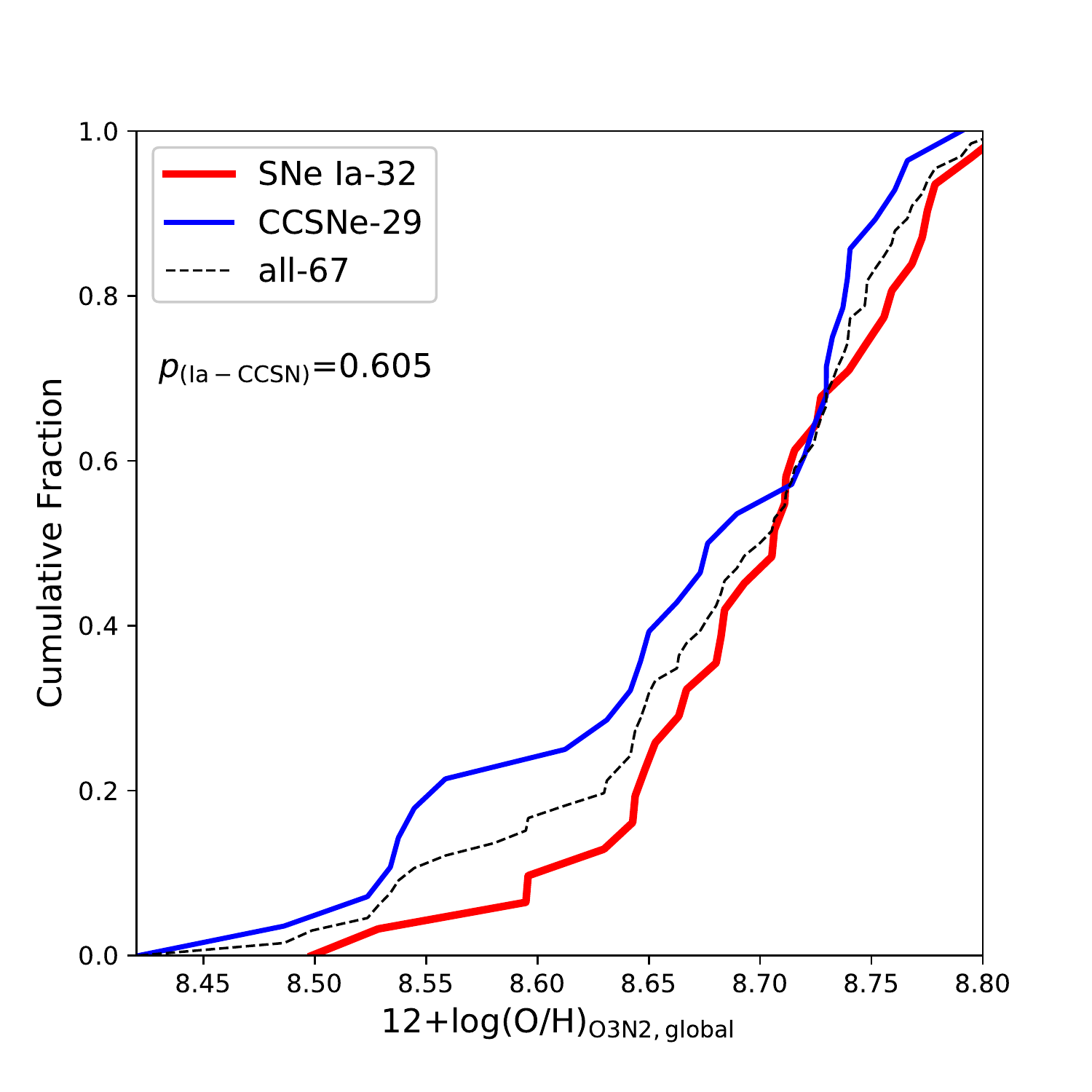}
\caption {The cumulative distributions of global gas-phase oxygen abundance for different types of SN host galaxies. }
\label{fig.oh-global}
\end{figure}

The cumulative distributions of global gas-phase oxygen abundance for different types of SN hosts are presented in Figure~\ref{fig.oh-global}. Here the gas-phase oxygen abundance is estimated using O3N2 method. From the K-S test, we can see that there is a high probability (p = 0.605) for SN Ia and CCSN hosts with the same distribution, which means the difference of global gas-phase oxygen abundance between SN Ia and CCSN hosts is small.

\begin{figure}
\centering
\includegraphics[angle=0,width=7.8cm]{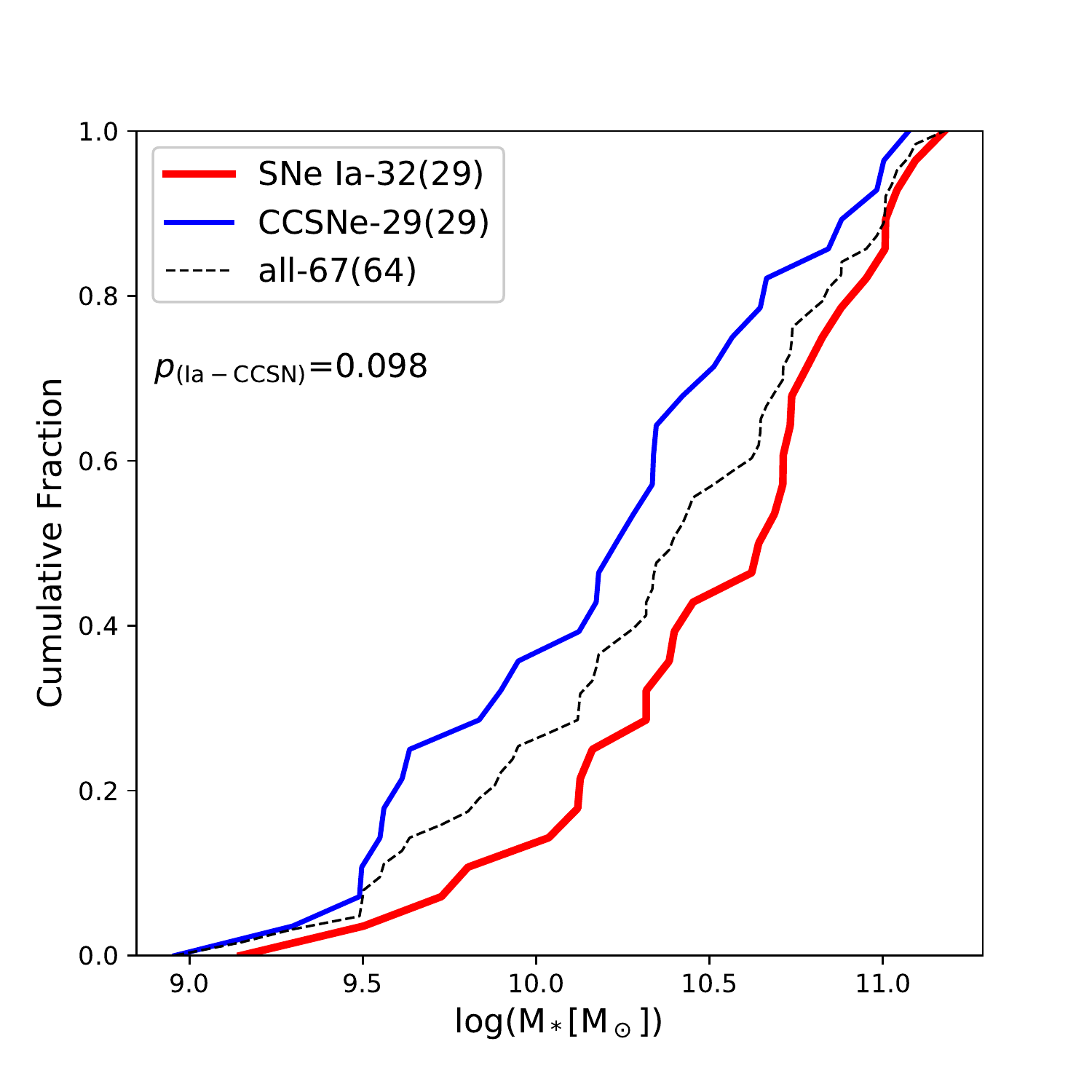}
\caption {The cumulative distributions of global stellar mass for different types of SN host galaxies. }
\label{fig.mass-global}
\end{figure}

Here we adopt the stellar mass from NSA, which can provide the stellar mass of the whole galaxy. It is estimated based on the photometry image of the whole galaxy. The cumulative distributions of global stellar mass for different types of SN hosts are shown in Figure~\ref{fig.mass-global}. From this figure we can see that the stellar mass of our sample galaxies range from $10^9$ to $10^{11.5}$ $M_\odot$. There are only less than 20\% of SNe Ia and 30\% of CCSNe found to explode in galaxies with stellar mass lower than $10^{10}$ $M_\odot$. This can be explained by the way of searching nearby SNe, which are often observed by targeting on the bright massive galaxies. Thus, galaxies with lower stellar mass is unrepresentative for nearby SN sample\citep{2010ApJ...715..743K,2009ApJ...707.1449N,2010ApJ...721..777A,ga14}. According to the p-value of K-S test, there is a significant difference of stellar mass between SN Ia hosts and CCSN hosts. From this figure and Table~\ref{table.med}, SN Ia hosts are more massive by $\sim$ 0.3 dex than CCSN hosts on average.    

\begin{figure}
\centering
\includegraphics[angle=0,width=8.8cm]{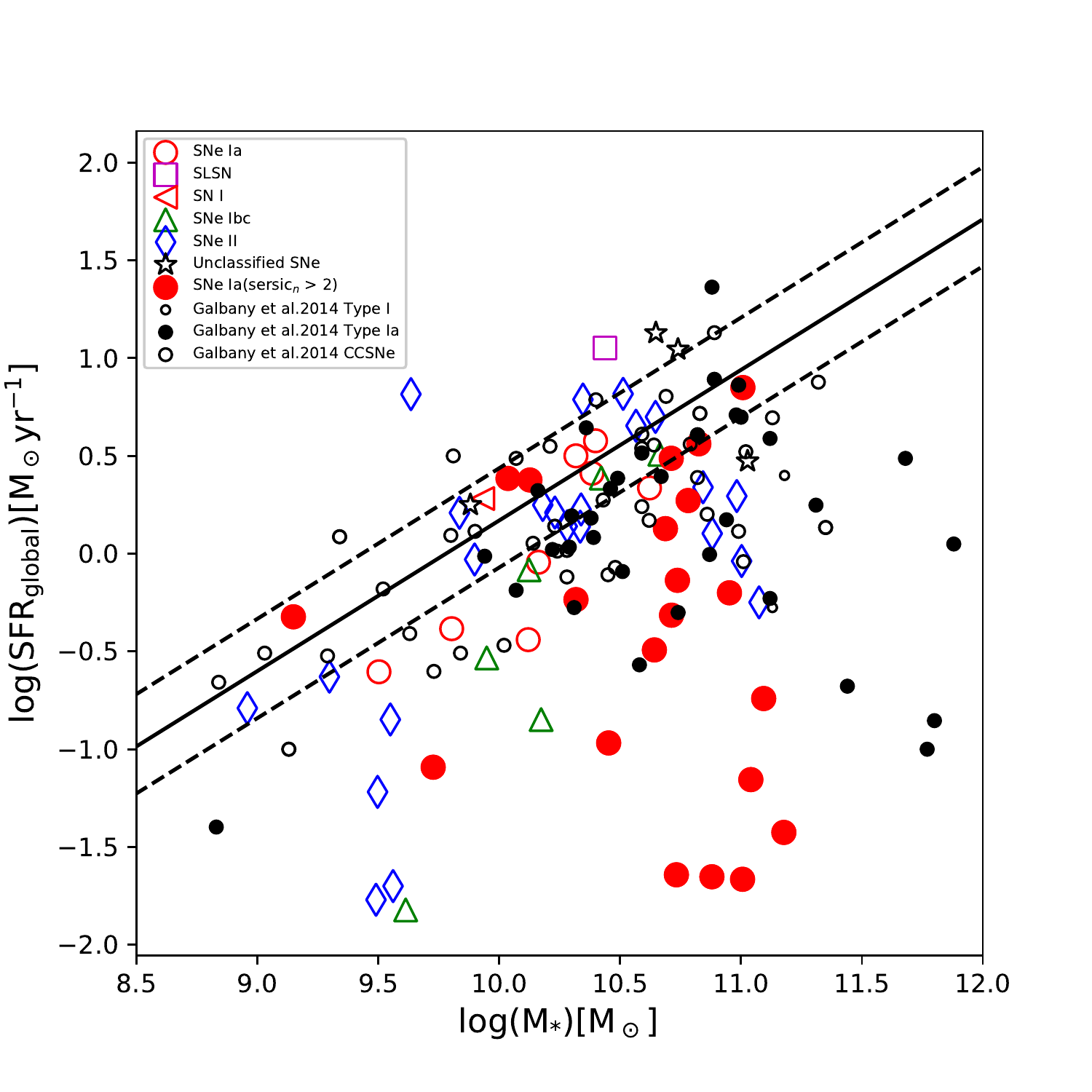}
\caption {The relation between global stellar mass and SFR for different types of SN host galaxies.}
\label{fig.sfr-m-67}
\end{figure}

We present the relation between stellar mass and global SFR of our sample galaxies and that for sample galaxies from \citet{ga14} in Figure~\ref{fig.sfr-m-67}. 
We should note that the stellar mass adopted for our sample galaxies is that of the whole galaxies, while the global SFR is for the part of galaxies within 1.5 $R_e$ of galaxies. 
We can see that most of our sample galaxies locate close to the relation defined by \citet{el07} and \citet{br04}, which studied star forming galaxies with redshift $\sim$ 0 from SDSS data. There are some galaxies with higher stellar mass and lower SFR deviate largely from the locus of \citet{el07}. Most of these galaxies are the hosts of SN Ia. On the whole, the relation for our sample galaxies is consistent with that of some literatures \citep{br04,el07,ga14,2019RAA....19..121Z}.   

\begin{figure}
\centering
\includegraphics[angle=0,width=8.8cm]{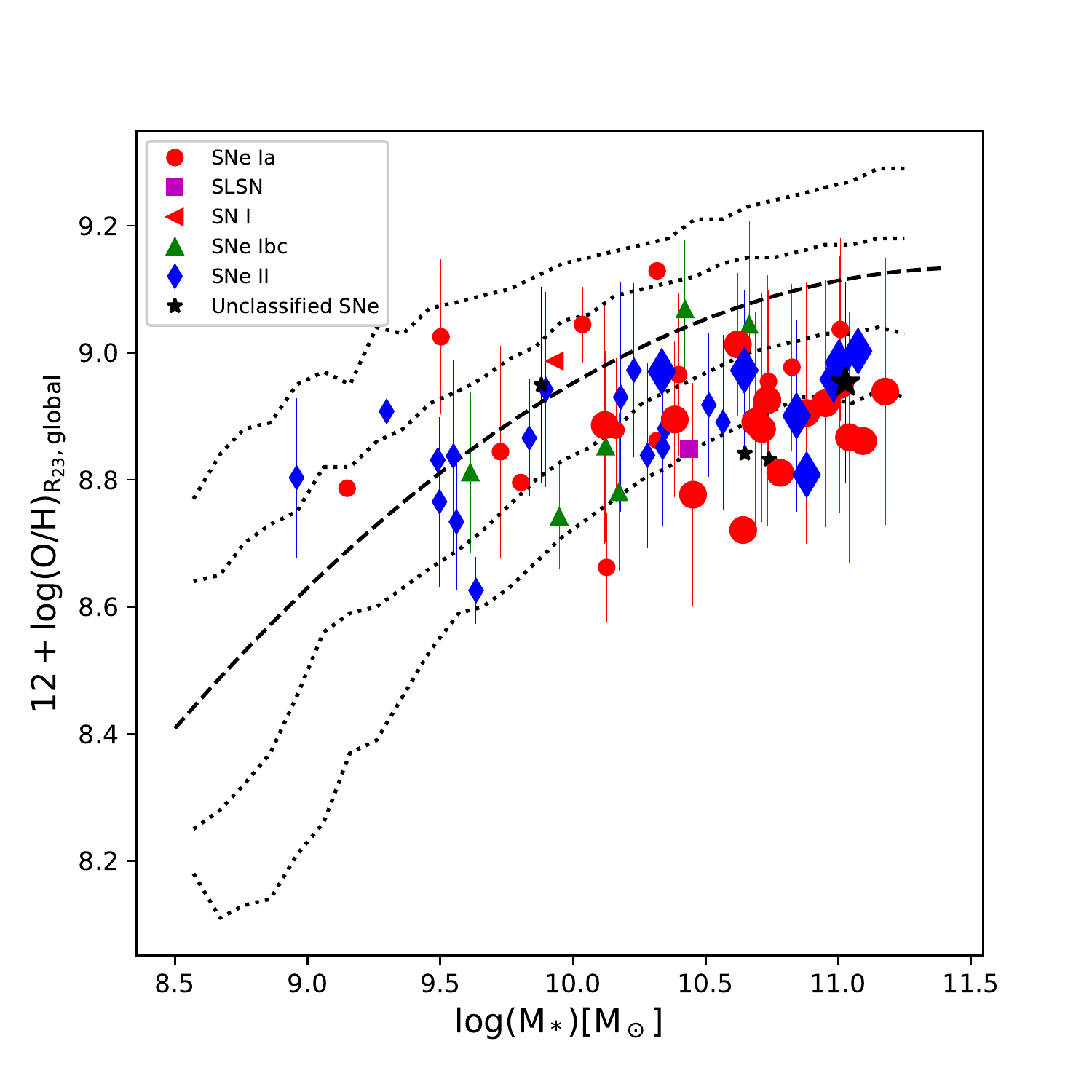}
\caption {The relation between global stellar mass and gas-phase oxygen abundance estimated by $R_{23}$ for different types of SN host galaxies.}
\label{fig.oh-m-67}
\end{figure}

We present the distributions of stellar mass of the whole galaxies and global gas-phase oxygen abundance and compare them with those of \citet{tr04} in Figure~\ref{fig.oh-m-67}. Here we choose gas-phase oxygen abundance estimated using $R_{23}$ method because the measurements in the relation in \citet{tr04} are estimated using $R_{23}$ method. In this figure, the larger symbols represent galaxies with AGNs in the center. From this figure, we can see that most of our sample galaxies locate within or near the region of 95\% level determined by \citet{tr04}. Some massive galaxies deviate slightly from the 95\% level and they have lower global gas-phase oxygen abundance. These massive galaxies mostly host SNe Ia and there are AGNs in the centers of galaxies. The relation between stellar mass and gas-phase oxygen abundance in \citet{tr04} is only for star-forming galaxies in SDSS. However, for some of our SN host galaxies, especially for some SNe Ia, their host galaxies are ellipticals. In our sample, there are AGNs in the central regions for 25 SN hosts. When estimating the gas-phase oxygen abundance, we remove out the central regions for galaxies with AGNs in the centers. Therefore, the global gas-phase oxygen will be lower and locate below the lines of \citet{tr04}. 

Also, the deviation may be caused by the systematically difference between the way of stellar mass and gas-phase oxygen abundance estimation. In our work, we estimate the gas-phase oxygen abundance by calculating the median value of the spaxels with emission line fluxes in 1.5 Re of galaxies. However, in \citet{tr04}, they use SDSS single-fiber spectra to obtain the gas-phase oxygen abundance. Therefore, our result may be systematically lower than that in \citet{tr04}. The stellar mass in our work is derived from NSA, which is based on the photometry image of the whole galaxy and can hence provide us the mass of the whole galaxy. While the mass from \citet{tr04} is based on the method in \citet{ka03b}, which are measured based on a single fiber spectrum of the nucleus and then scaled to the whole galaxy. 

\begin{figure}
\centering
\includegraphics[angle=0,width=8.8cm]{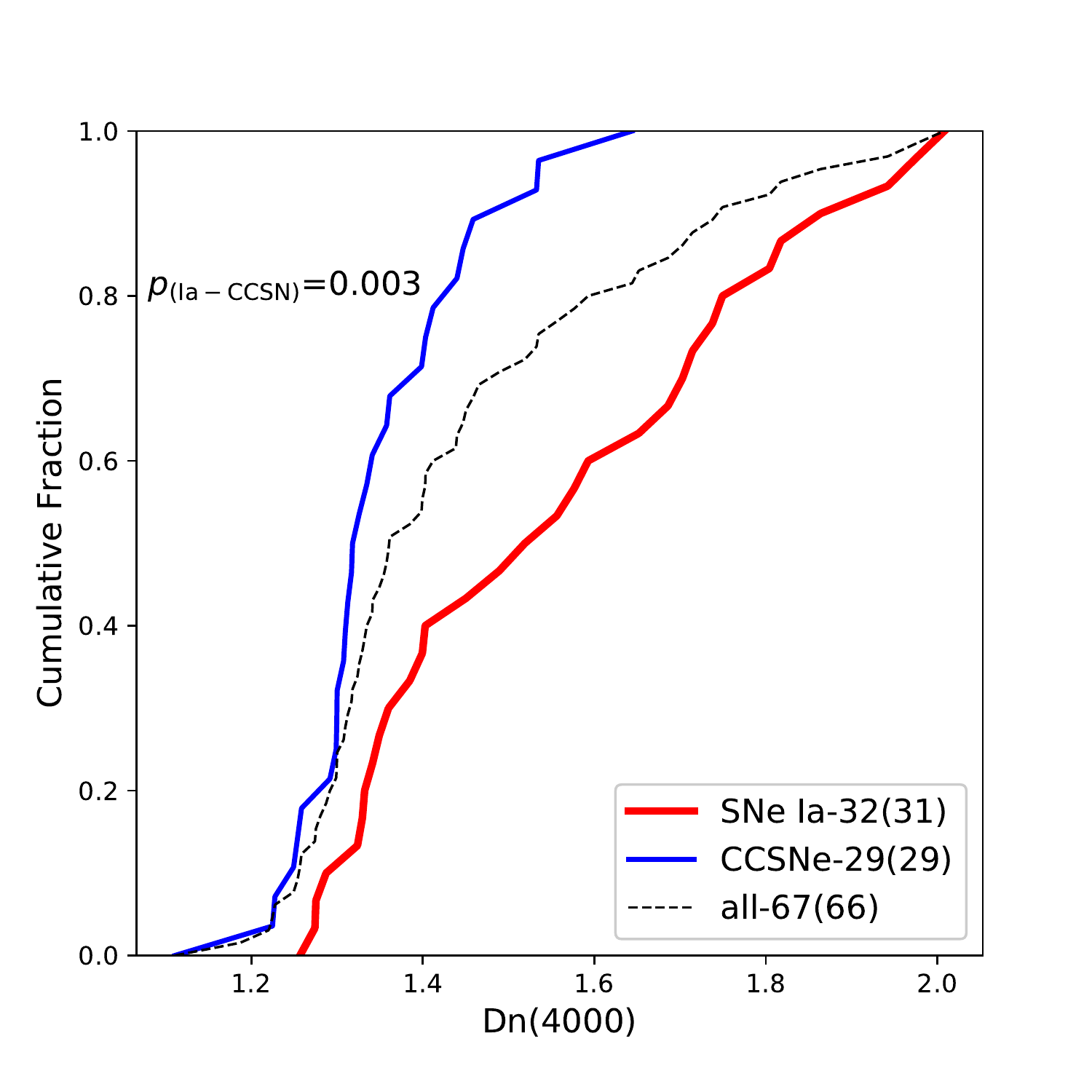}
\caption {The cumulative distributions of global Dn(4000) for different types of SN host galaxies.}
\label{fig.cdf-d4000-67}
\end{figure}

Dn(4000) is well known as an indicator of stellar population age \citep{br83,bal99,ka03a}. In our work, the global Dn(4000) and $H\delta_A$, which will be described below, are estimated by calculating the median value of spaxels within 1.5 $R_e$ of galaxies. We present the cumulative distributions of the global Dn(4000) of different types of SN host galaxies in Figure~\ref{fig.cdf-d4000-67}. From this figure we can see that there is a significant difference of global Dn(4000) between SN Ia and CCSN hosts. On average, the Dn(4000) of SN Ia hosts is larger than that of CCSN hosts, which means that the stellar population of SN Ia hosts is older than CCSN hosts. 

\begin{figure}
\centering
\includegraphics[angle=0,width=8.8cm]{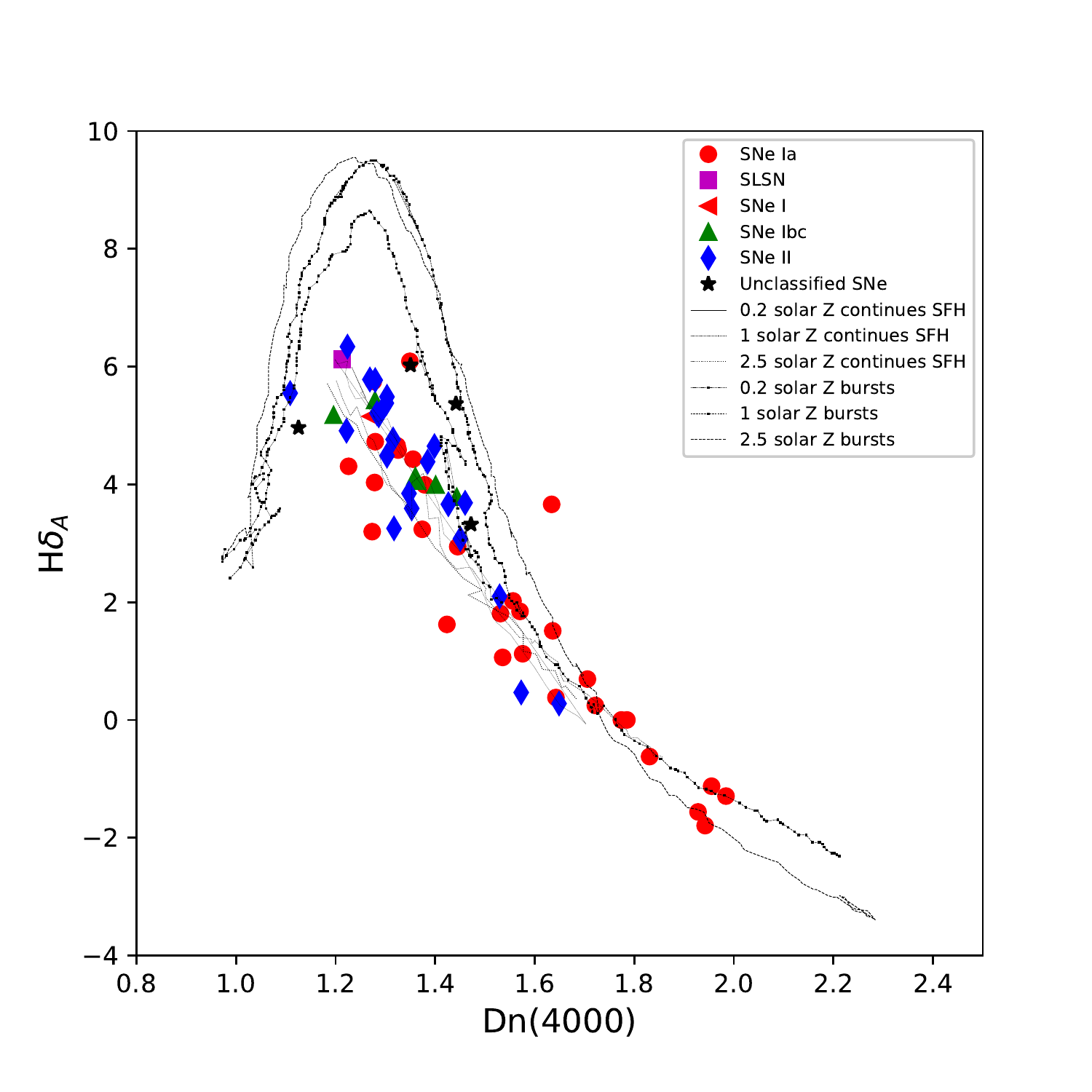}
\caption {The distribution of the global Dn(4000) and $\rm H\delta_A$ for our SN galaxies. The dashed  lines are the relations between Dn(4000) and $\rm H\delta_A$ for 20$\%$ solar, solar and 2.5 times solar metallicity bursts, respectively. The dotted lines are the relations between Dn(4000) and $\rm H\delta_A$ for 20$\%$ solar, solar and 2.5 solar metallicity continuous star formation histories. These lines are taken from \cite{ka03b}. }
\label{fig.d4000-hdelta-67}
\end{figure}

$H\delta_A$ can also trace the stellar population of galaxies\citep{wo97}. 
In Figure~\ref{fig.d4000-hdelta-67}, we present the relations between Dn(4000) and $H\delta_A$ for our sample galaxies and compare them with \citet{ka03b}, which studied pure burst star formation histories and continuous star formation histories with different metallicity. From Figure~\ref{fig.d4000-hdelta-67}, we can see that the relations for our sample galaxies is well consistent with \citet{ka03b}. In this relation, the global values of Dn(4000) for SN Ia host galaxies are larger than CCSN host galaxies and the global values of $H\delta_A$ for SN Ia host galaxies are lower than CCSN host galaxies, which means that the stellar population of CCSN host galaxies is younger than SN Ia hosts. This result is consistent with the cumulative distribution of Dn(4000). 

\begin{table*}
\centering
\begin{tabular}{p{15mm}p{30mm}p{30mm}p{30mm}p{30mm}}
\hline
\multirow{2}{*}{SNe Type} &	\multicolumn{2}{c}{Global} & \multicolumn{2}{c}{Local} \\
 & median  & mean &   median  & mean   \\
\hline
  & \multicolumn{2}{c}{log($SFR$)[$M_\odot$ $yr^{-1}$]} & \multicolumn{2}{c}{log($\Sigma_{SFR}$)[$M_\odot$ $yr^{-1}$ $kpc^{-2}$]} \\	
SNe Ia & -0.24 $\pm$ 0.73 & -0.30 $\pm$ 0.73  & -2.15 $\pm$ 1.34 & -2.54 $\pm$ 1.34 \\
CCSNe & 0.14 $\pm$ 0.76 & -0.14 $\pm$ 0.76 & -2.17 $\pm$ 1.10 & -2.57 $\pm$ 1.10  \\
All & 0.12$\pm$ 0.77 & -0.13 $\pm$ 0.77 & -2.10 $\pm$ 1.19 & -2.49 $\pm$ 1.19 \\
\hline
\multicolumn{5}{c}{log($M_*$)[$M_\odot$]} \\
SNe Ia & 10.63 $\pm$ 0.49 & 10.47 $\pm$ 0.49 &...&...\\
CCSNe & 10.26 $\pm$ 0.56 & 10.18 $\pm$ 0.56 & ...& ...\\
All & 10.39 $\pm$ 0.54 & 10.34 $\pm$ 0.54 &...&...\\
\hline
\multicolumn{5}{c}{$EW_{H\alpha}[\AA]$} \\
SNe Ia & 5.78 $\pm$ 8.84 & 8.14 $\pm$ 8.84 & 4.08 $\pm$ 11.18 & 9.33 $\pm$ 11.18 \\
CCSNe & 14.65 $\pm$ 16.66 & 16.65 $\pm$ 16.66 & 21.23 $\pm$ 13.97 & 23.60 $\pm$ 13.97 \\
All & 10.20 $\pm$ 13.87 & 13.11 $\pm$ 13.87 & 12.60 $\pm$ 14.80 & 16.67 $\pm$ 14.80\\
\hline
\multicolumn{5}{c}{$H\delta_A[\AA]$} \\
SNe Ia & 1.84 $\pm$ 2.28& 2.15 $\pm$ 2.28 & 1.83 $\pm$ 2.46 & 2.05 $\pm$ 2.46\\
CCSNe &4.38 $\pm$ 1.56  & 4.10 $\pm$ 1.56 & 4.61 $\pm$ 1.33 & 4.53 $\pm$ 1.33\\
All & 3.92 $\pm$ 2.18 & 3.32 $\pm$ 1.98 & 4.06 $\pm$ 2.39 & 3.46 $\pm$ 2.39\\
\hline
\multicolumn{5}{c}{$Dn(4000)$} \\
SNe Ia & 1.56 $\pm$ 0.28 &1.59 $\pm$ 0.28 & 1.50 $\pm$ 0.41 & 1.48 $\pm$ 0.41\\
CCSNe &1.32 $\pm$ 0.21 &1.32 $\pm$ 0.21 & 1.32 $\pm$ 0.12& 1.34 $\pm$ 0.12\\
All & 1.32 $\pm$ 0.28 &1.44 $\pm$ 0.28 & 1.37 $\pm$ 0.30 & 1.41 $\pm$ 0.30\\
\hline
\hline
\multicolumn{5}{c}{12+log(O/H)} \\	
\multicolumn{5}{c}{O3N2} \\	
SNe Ia & 8.71 $\pm$ 0.07 & 8.69 $\pm$ 0.07 & 8.72 $\pm$ 0.10 & 8.71 $\pm$ 0.10 \\
CCSNe & 8.68 $\pm$ 0.09 & 8.66 $\pm$ 0.09 & 8.74 $\pm$ 0.11 & 8.68 $\pm$ 0.11 \\
All & 8.70 $\pm$ 0.08 & 8.68 $\pm$ 0.08 & 8.72 $\pm$ 0.10 & 8.70 $\pm$ 0.10 \\
\hline
\multicolumn{5}{c}{12+log(O/H)} \\	
\multicolumn{5}{c}{N2O2} \\	
SNe Ia & 8.96 $\pm$ 0.09 & 8.96 $\pm$ 0.09 & 8.97 $\pm$ 0.11 & 8.97 $\pm$ 0.11 \\
CCSNe & 8.91 $\pm$ 0.11 & 8.90 $\pm$ 0.11 & 8.94 $\pm$ 0.12 & 8.92 $\pm$ 0.12 \\
All & 8.95 $\pm$ 0.10 & 8.93 $\pm$ 0.10 & 8.96 $\pm$ 0.11 & 8.95 $\pm$ 0.11 \\
\hline
\multicolumn{5}{c}{12+log(O/H)} \\	
\multicolumn{5}{c}{R23} \\	
SNe Ia & 8.91 $\pm$ 0.07 & 8.91 $\pm$ 0.07 & 8.88 $\pm$ 0.15 & 8.90 $\pm$ 0.15\\
CCSNe & 8.88 $\pm$ 0.10 & 8.88 $\pm$ 0.10  & 8.92 $\pm$ 0.13 & 8.91 $\pm$ 0.13 \\
All & 8.89 $\pm$ 0.10 & 8.89 $\pm$ 0.10 & 8.92 $\pm$ 0.14 & 8.90 $\pm$ 0.14 \\
\hline
\end{tabular}
\caption{The median and mean values with standard deviations of the local and global properties for host galaxies of Type Ia, CCSNe and all the sample galaxies. }
\label{table.med}
\end{table*}

\subsection{The difference of local properties between different types of SNe host galaxies}

\begin{figure}
\centering
\includegraphics[angle=0,width=8.8cm]{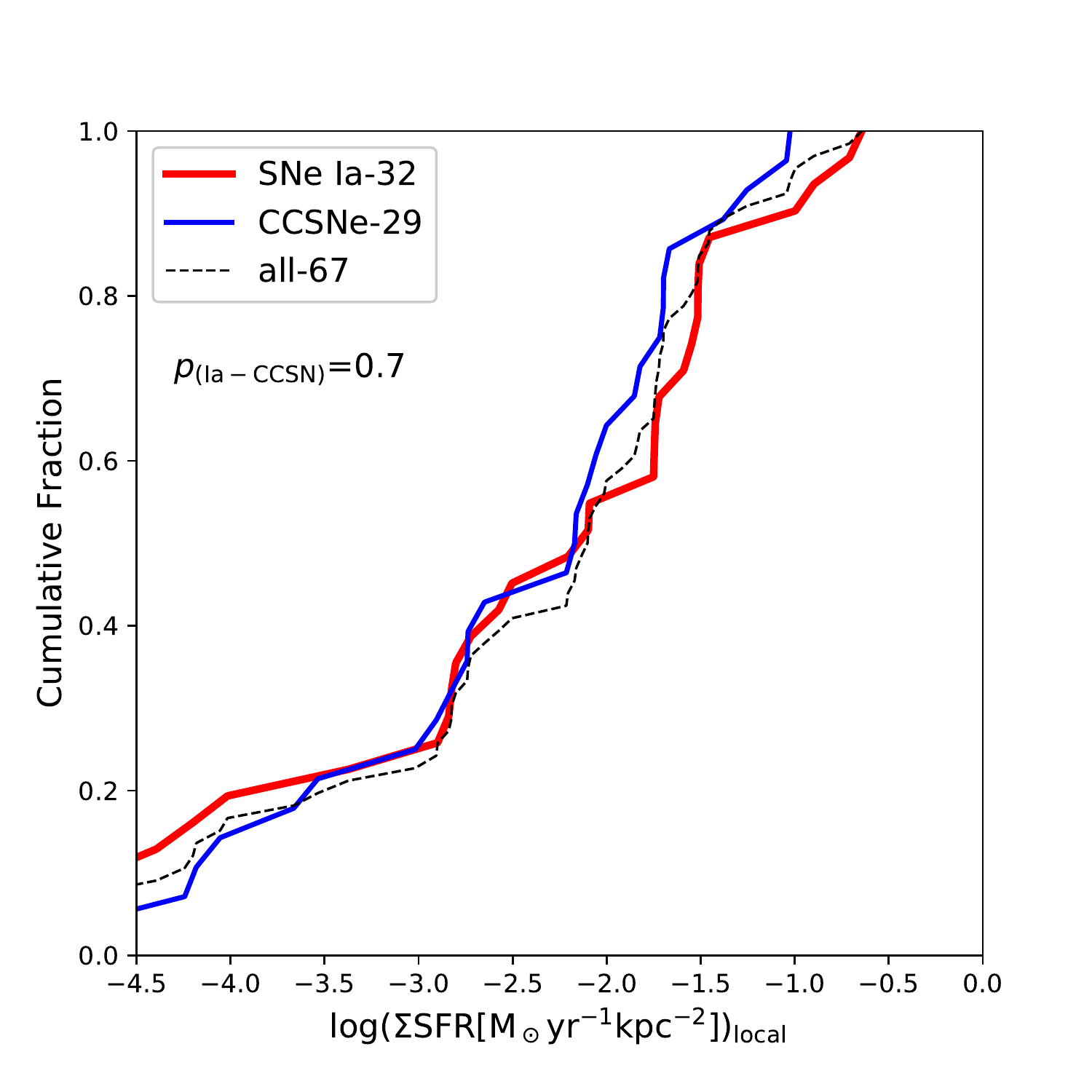}
\caption {The cumulative distributions of local $\Sigma \rm SFR$ for different types of SN host galaxies at SN explosion sites.}
\label{fig.cdf-sfr-local-67}
\end{figure}

We estimate the local properties of SN explosion sites using emission line fluxes of useful spaxels in the circle region with a radius of 2.5 $arcsecond$. Figure~\ref{fig.cdf-sfr-local-67} shows the cumulative distributions of local star formation density ($\Sigma \rm SFR$) at the SN explosion sites. We obtain a high p-value of K-S test (p = 0.70), which means that there is no significant differences between the local $\Sigma \rm SFR$ of SN Ia explosion sites and that of CCSN. 

\begin{figure}
\centering
\includegraphics[angle=0,width=8.8cm]{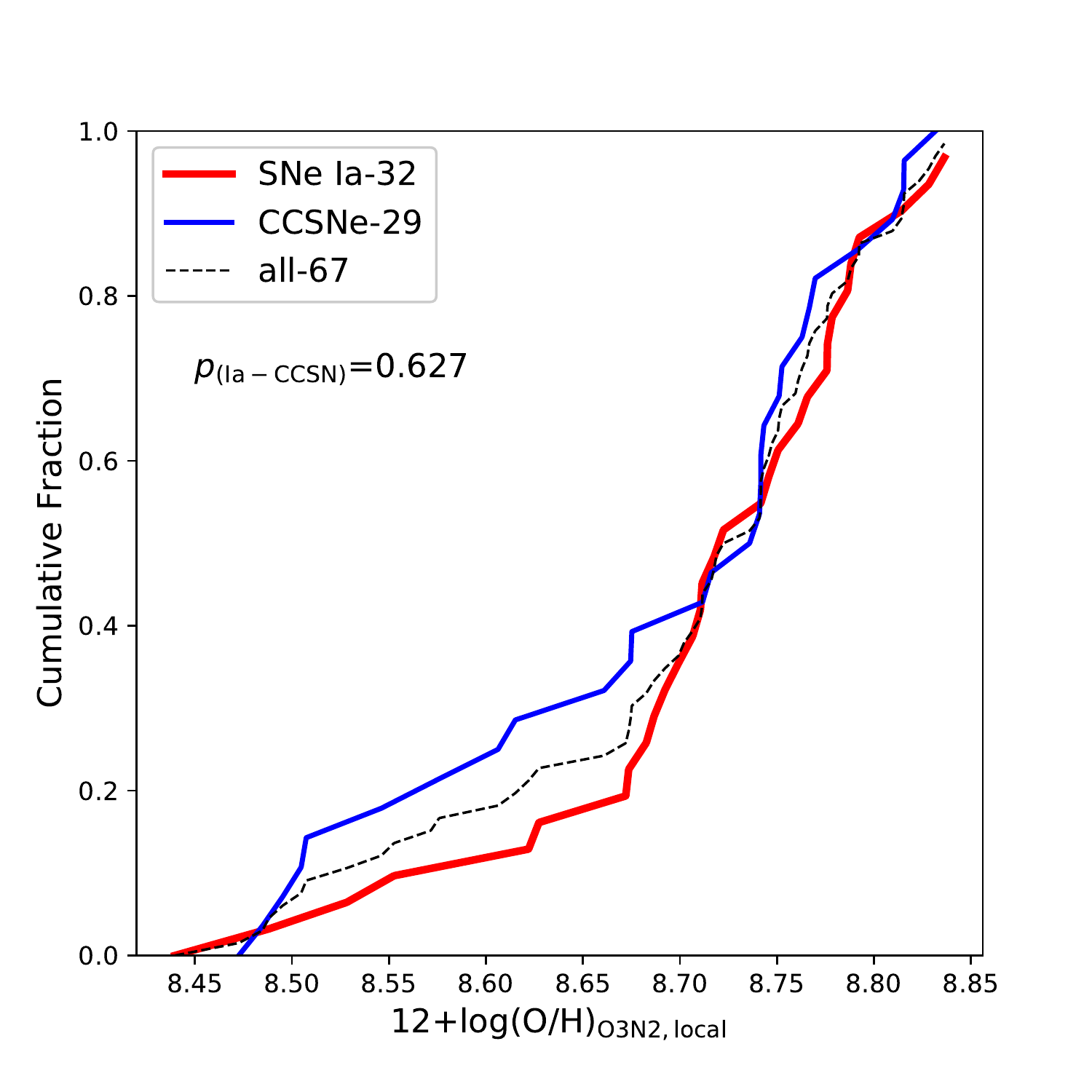}
\caption {The cumulative distributions of local gas-phase oxygen abundance for different types of SN host galaxies at SN explosion sites.}
\label{fig.cdf-oh-local-67}
\end{figure}

In figure~\ref{fig.cdf-oh-local-67}, the cumulative distributions of local gas-phase oxygen abundance for different types of SN explosion sites are presented. From this figure, we can see that the local gas-phase oxygen abundance of SN Ia explosion sites is slightly higher than that of CCSNe on average. In general, there is no significant differences between the local gas-phase oxygen abundance at the explosion sites for different types of SNe.   

\section{Discussion}
\label{discussion}

\subsection{Comparisons with the measurements in DR13 }



\begin{figure}
\centering
\includegraphics[angle=0,width=8.8cm]{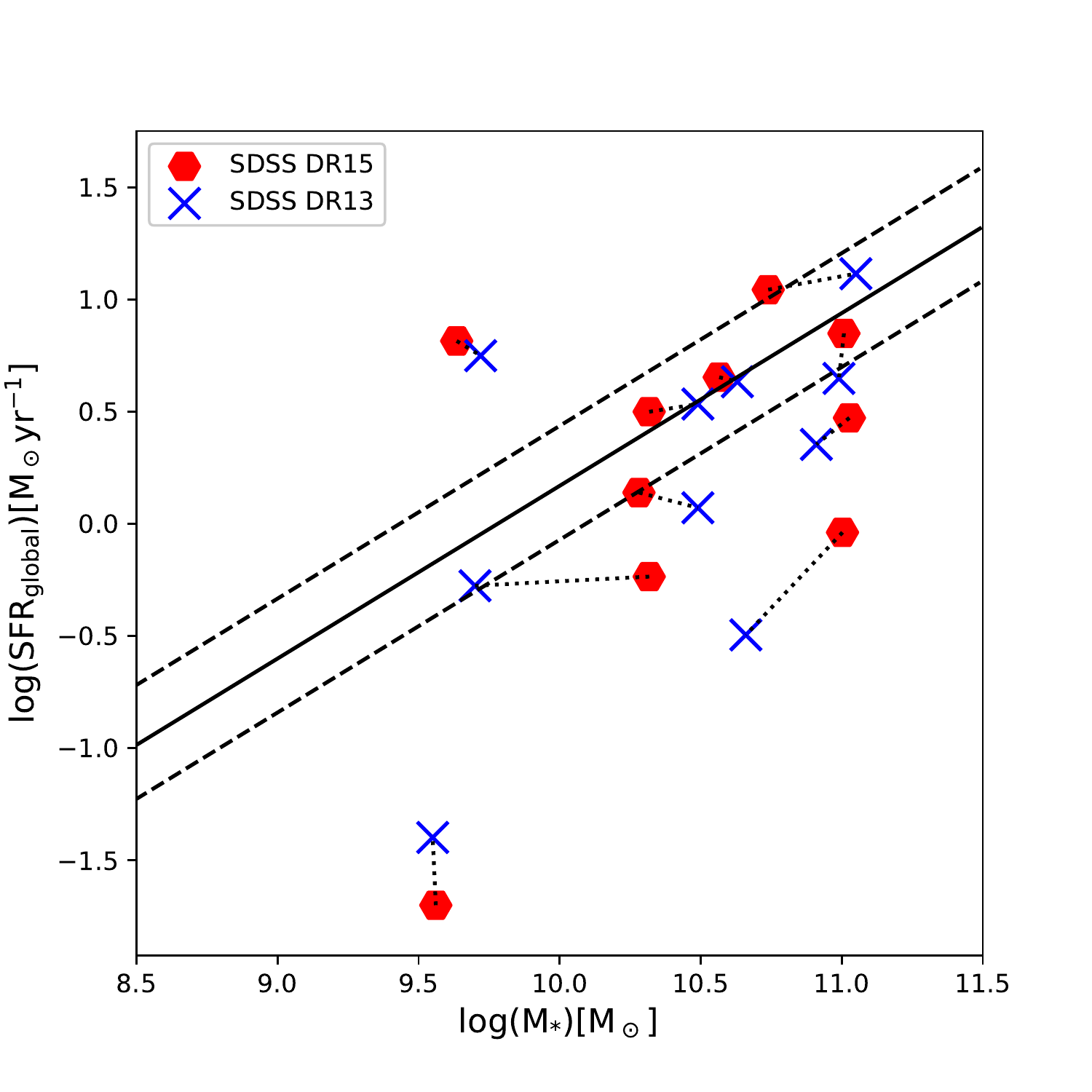}
\caption {The relations of global SFR and stellar mass for the same sample galaxies for DR13 in \citet{2019RAA....19..121Z} and DR15 data in this paper.}
\label{fig.m-sfr_dr13_dr15}
\end{figure}

We compare the distributions of stellar mass and star formation rate of galaxies in DR15 in this work with those in DR13 studied by \citet{2019RAA....19..121Z} in Figure~\ref{fig.m-sfr_dr13_dr15} for the same sample objects. The red hexagons and blue pluses represent measurements in DR15 and DR13, respectively. The stellar masses in DR15 are derived from NSA and those in DR13 are estimated using {\tt STARLIGHT} code. From this figure, we can see that the relations between stellar mass and star formation rate for DR15 data are well consistent with those for DR13 data in \citet{2019RAA....19..121Z}.


\begin{figure}
\centering
\includegraphics[angle=0,width=8.8cm]{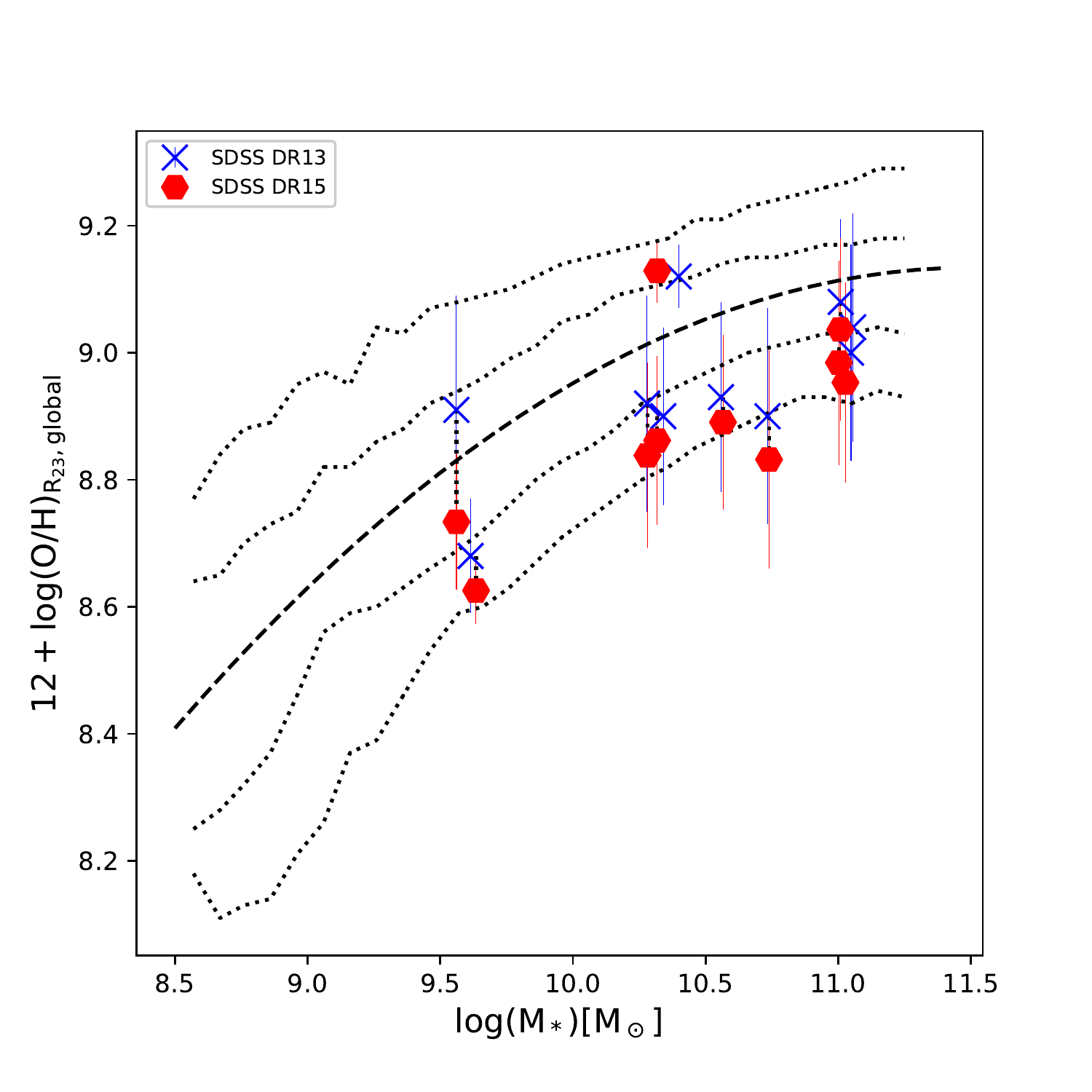}
\caption {The relations of global gas-phase oxygen abundance and stellar mass for the same sample galaxies for DR13 in \citet{2019RAA....19..121Z} and DR15 data in this paper.}
\label{fig.m-lgoh_dr13_dr15}
\end{figure}

In Figure~\ref{fig.m-lgoh_dr13_dr15}, the relations between stellar mass and  global gas-phase oxygen abundance estimated by $R_{23}$ method for the same sample galaxies in both \citet{2019RAA....19..121Z} and this work are presented. We can see that the difference between the present measurements in DR15 and those in DR13 is small. All of the same sample galaxies locate within or nearby the lines that enclose 95\% of the data from \cite{tr04}.


\subsection{The difference between local and global gas-phase oxygen abundance for our sample galaxies}

The specific local properties of SN explosion sites are important to derive the properties of progenitors. However, for some SNe exploding in distant galaxies, whose spectral can often be obtained with long-slit or fixed aperture fibers, only global properties of galaxies can be measured. Therefore, we try to study the correlation between local and global properties. According to the correlation, we can estimate the local properties of SN explosion sites through the global properties. Due to our sample limitation that there is a lack in low mass host galaxies, we can use IFU data to study the difference between local and global properties for now. 

Here we should note that the progenitors of SN Ia may be very old stars \citep{2011MNRAS.412.1508M} and the local properties at their recent locations may be different from those of the regions where they originally formed, which is because that they may have migrated from birth place.

\begin{figure}
\centering
\includegraphics[angle=0,width=8.8cm]{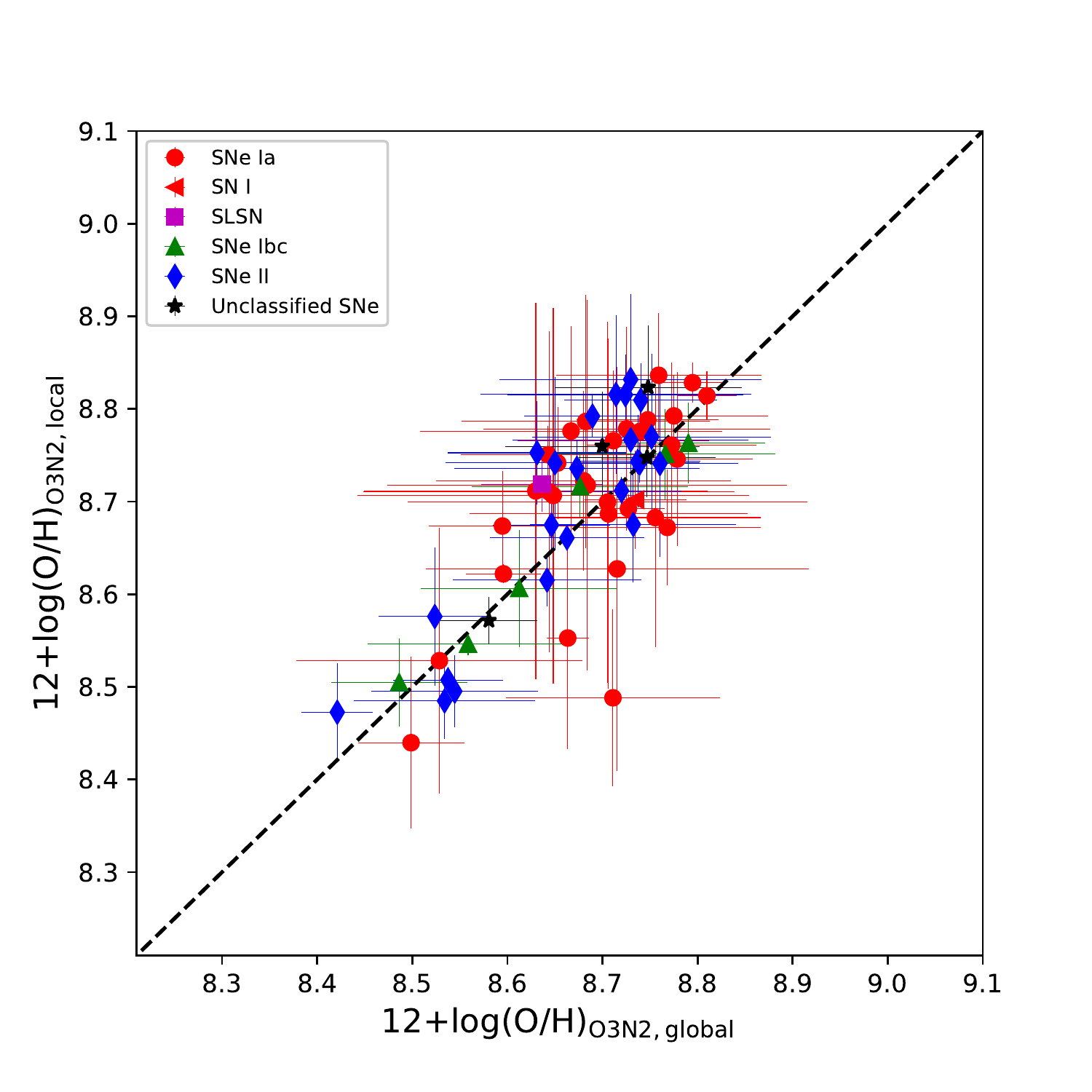}
\caption {The difference between local and global gas-phase oxygen abundance for  different types of SN host galaxies.}
\label{fig.delta-oh-67}
\end{figure}

The difference of local and global gas-phase oxygen abundance are shown in Figure~\ref{fig.delta-oh-67}. From this figure we can see that most of our sample galaxies locate close to the dashed line, which represents the ratio of local and global gas-phase oxygen abundance is 1:1. This is consistent with \citet{ga16b} and \citet{2019RAA....19..121Z}. There are some galaxies deviate slightly from the dashed line and most of these galaxies host SNe Ia.


\subsection{The differences of measurements between different types of SN host galaxies}

Through the cumulative distributions of global SFR, we can see that for our sample galaxies, CCSNe tend to explode in hosts with higher SFR than SNe Ia. As \citet{ke98} and \citet{2009ApJ...691..115G} pointed out that the massive stars, which are thought to be the progenitors of CCSNe, explode near the sites where they born and their lifetimes are shorter than less massive stars. The explosion of massive stars could cause the decrease of $H\alpha$ emission intensity and peculiar motions of less massive stars around them \citep{2011MNRAS.414.3501E}. Then the less massive stars would explode to SNe Ia in a lower $H\alpha$ emission environment that are further away from the centers of star formation region. 
SNe Ia can be observed both in ellipticals and in spiral galaxies, which contain old and young stellar populations \citep{2012A&A...540A..11K}. Therefore, the SFR of SN Ia hosts is lower than that of CCSN hosts on average.

In our sample, SN Ia hosts are more massive than CCSN hosts on average. The mean logarithm stellar mass for SN Ia host galaxies is 10.47 $\pm$ 0.49 and 10.18 $\pm$ 0.56 for CCSN hosts. The difference of the stellar mass between SN Ia and CCSN hosts is $\sim$ 0.3 dex on average. This indicates that the ratio of CCSNe to SNe Ia will decrease with increasing stellar mass of galaxies, which is consistent with \citet{ga14}. 
As \citet{ga14} pointed out that this can be explained by the different delay-time distribution (DTD) for SNe Ia and CCSNe. There are about half of the progenitors of SNe Ia older than 1 Gyr \citep{2010ApJ...722.1879M,2011MNRAS.412.1508M}, while CCSNe explode within about 40 Myr after the star formation.      
There is a positive relation between SFR and stellar mass within $10^{11}$ $M\odot$ \citep{el07}. Compared with less massive galaxies, there is a larger fraction of old stellar populations in more massive galaxies. Therefore, the ratio of CCSNe to SNe Ia will decrease with the stellar mass of galaxies increasing.

The difference of both global and local gas-phase oxygen abundance between SN Ia and CCSN hosts is small for our sample. The average global gas-phase oxygen abundance estimated by O3N2 method for SN Ia and CCSN hosts is 8.69 and 8.66, respectively. The mean local gas-phase oxygen abundance for both types is 8.71 and 8.68, respectively. The differences both for global and local are significantly smaller than 1 $\sigma$. 

\section{Conclusion}
\label{conclusion}
In this work, we present the local and global properties, including SFR, gas-phase oxygen abundance, galaxy mass and Dn(4000) etc., for 67 different types of SN host galaxies selected from SDSS DR15 within the FoV of MaNGA. 
There are 32 Type Ia, 29 CCSNe, 1 SLSN, 1 Type I and 4 unclassified type of SNe in our sample. 
Using spatially resolved IFS of MaNGA, we could derive the cumulative distribution of local environment of SN explosion sites and the global properties of different types of SN hosts. Also, we compare the local with global gas-phase oxygen abundance to derive the differences between them. Like our first paper of a series in \cite{2019RAA....19..121Z}, our sample from MaNGA has higher redshifts with the median of $\sim$ 0.03, which can be used to analyze more distant SN host galaxies. We would like to point out that in our work, due to the limited sample size, we couldn't remove the mass dependence for different types of SN host galaxies, which is likely the true driver of the trends for the properties presented in this work. Our results are concluded as follows.

On average, the fact that global SFR of SN Ia host galaxies is lower than that of CCSN hosts is consistent with the results of EW(H$\alpha$), which represents that the star formation activity for CCSN hosts is stronger than SN Ia hosts. 
The mean global stellar mass of SN Ia host galaxies is $\sim$ 0.3 dex higher than that of CCSN hosts, which indicates that with the increasing stellar mass of galaxies, the number ratio of SNe Ia to CCSNe will increase. 
The global gas-phase oxygen abundance between different types of SN host galaxies is almost similar. According to the Dn(4000) distribution of different types of SN hosts, the stellar population age for SN Ia host galaxies is older than that of CCSN hosts on average. 
  
There is no significant difference of local gas-phase oxygen abundance and local star formation rate density for SN Ia and CCSN explosion sites in their host galaxies. 
For most of our sample galaxies, the difference between global and local gas-phase oxygen abundance is small. While for some SN Ia host galaxies, the difference is larger.  

There will be a larger sample of SN host galaxies in MaNGA survey, which has observed  $\sim$ 4600 galaxies for now, and there will be 10,000 galaxies in the end of this survey. At that time, we could obtain more and more SN hosts and make a more detailed and fair comparison between the explosion environments and progenitors of different types of SNe after removing mass dependence . 

\begin{acknowledgements}
We appreciate the referee who provided very constructive and
helpful comments and suggestions, which helped to improve very
well our work. 
This work was supported by the National Science Foundation of China (Grant No.11733006 to HW, 11903046 to JG and U1631105 to WB), and by the Beijing Municipal Natural Science Foundation (No. 1204038 to JG).


Funding for the Sloan Digital Sky Survey IV has been provided by the Alfred P. Sloan Foundation, the U.S. Department of Energy Office of Science, and the Participating Institutions. SDSS acknowledges support and resources from the Center for High-Performance Computing at the University of Utah. The SDSS web site is www.sdss.org.

SDSS is managed by the Astrophysical Research Consortium for the Participating Institutions of the SDSS Collaboration including the Brazilian Participation Group, the Carnegie Institution for Science, Carnegie Mellon University, the Chilean Participation Group, the French Participation Group, Harvard-Smithsonian Center for Astrophysics, Instituto de Astrofísica de Canarias, The Johns Hopkins University, Kavli Institute for the Physics and Mathematics of the Universe (IPMU) / University of Tokyo, the Korean Participation Group, Lawrence Berkeley National Laboratory, Leibniz Institut für Astrophysik Potsdam (AIP), Max-Planck-Institut für Astronomie (MPIA Heidelberg), Max-Planck-Institut für Astrophysik (MPA Garching), Max-Planck-Institut für Extraterrestrische Physik (MPE), National Astronomical Observatories of China, New Mexico State University, New York University, University of Notre Dame, Observatório Nacional / MCTI, The Ohio State University, Pennsylvania State University, Shanghai Astronomical Observatory, United Kingdom Participation Group, Universidad Nacional Autónoma de México, University of Arizona, University of Colorado Boulder, University of Oxford, University of Portsmouth, University of Utah, University of Virginia, University of Washington, University of Wisconsin, Vanderbilt University, and Yale University.
\end{acknowledgements}

\label{lastpage}
\end{document}